\pdfoutput=1

\documentclass[9pt,twocolumn]{IEEEtran}

\usepackage{graphicx}
\usepackage{epstopdf}
\usepackage{amsmath}
\usepackage{bbold}
\usepackage{enumerate}
\usepackage{cite}
\usepackage{color,xcolor}
\usepackage{amssymb}

\usepackage{algorithm}
\usepackage{algpseudocode}
\usepackage{subfigure}
\usepackage{epsfig}
\newtheorem{theorem}{Theorem}{}
\newtheorem{proposition}{Proposition}{}
\newtheorem{lemma}{Lemma}{}
\newtheorem{assumption}{Assumption}
\newtheorem{corollary}{Corollary}{}
\newtheorem{remark}{Remark}{}
\newtheorem{proof}{proof}{}

\begin{document}

\title{A Distributed Optimization Scheme for State Estimation of Nonlinear Networks with Norm-bounded  Uncertainties}

\author{Peihu Duan, ~Qishao~Wang,
        ~Zhisheng~Duan, \textit{Senior Member, IEEE,} and ~Guanrong Chen, \textit{Life Fellow, IEEE}% <-this % stops a space
\thanks{This work was supported by the National Key R$\&$D Program of China under Grant  2018AAA0102703, the National Natural Science Foundation of China under Grant U1713223, and the Hong Kong Research Grants Council under the GRF Grant CityU11206320. \emph{(Corresponding author: Zhisheng Duan.)}}

\thanks{P. Duan and Z. Duan are with the State Key Laboratory for Turbulence and Complex Systems, Department of Mechanics and Engineering Science, College of
Engineering, Peking University, Beijing, 100871, China. E-mails: duanpeihu@pku.edu.cn (P. Duan), duanzs@pku.edu.cn (Z. Duan).}

\thanks{Q. Wang is with the Department of Mechanical and Engineering Science, Beihang University, Beijing, China. E-mail: wangqishao@buaa.edu.cn (Q. Wang).}

\thanks{G. Chen is with the Department of Electrical Engineering, City University of Hong Kong, Hong Kong SAR, China. E-mail: eegchen@cityu.edu.hk (G. Chen).} }

\maketitle

\begin{abstract}
This paper investigates the state estimation problem for a class of complex  networks, in which the dynamics of each node is subject to Gaussian noise, system uncertainties and nonlinearities. Based on a regularized least-squares approach, the estimation problem is reformulated as an optimization problem, solving for a solution in a distributed way by utilizing a decoupling technique. Then, based on this solution, a class of estimators is designed to handle the system dynamics and constraints. A novel feature of this design lies in the unified modeling of uncertainties and nonlinearities, the decoupling of nodes, and the construction of recursive approximate covariance matrices for the optimization problem. Furthermore, the feasibility of the proposed estimators and the boundedness of mean-square  estimation errors are ensured under a developed criterion, which is easier to check than  some typical estimation strategies including the linear matrix inequalities-based  and the variance-constrained ones.  Finally, the effectiveness of the theoretical results is verified by a numerical simulation.
\end{abstract}


\section{Introduction}\label{sec1}
Over the past two decades, complex networks have drawn increasing attention since they can model many physical systems, such as cooperative unmanned aerial vehicles, networked manipulators and multi-satellite systems \cite{Arenas2008,wen2019pinning}.  One significant research issue is to estimate the system states of nodes in complex networks by utilizing system models and measurements \cite{duanpeihu2019,SHI2016275,duanpeihu2020,shen2011bounded}. The difference between the state estimation for a single system and that for a complex network lies in the coupling features between nodes, consequently the latter is much more complicated than the former. Moreover, there always exist disturbances, nonlinearities and uncertainties in system models, thus how to design a robust estimator for such a complex network is a great challenge.
Generally, the existing robust estimation strategies can be divided into two categories: the linear matrix inequalities (LMIs)-based approach and the variance-constrained approach.

Derived from the $H_{\infty}$ filtering for a single uncertain system, a large number of LMIs-based techniques have been developed for complex networks with various constraints, achieving an $H_{\infty}$ estimation performance  \cite{Ding2012H,shen2013H,wang2016anevent,Yong2017asy,Wu2018state,dong2018variance}.
Earlier, Ding et al. \cite{Ding2012H} proposed an estimator for complex networks with randomly occurring sensor saturations, where the exponential mean-square stability of the estimation was ensured under an augmented LMIs-based condition. Then, to reduce the data transmission frequency, Wang et al.  \cite{wang2016anevent}  embedded an event-triggered algorithm into the estimator for complex networks in the presence of mixed delays and Gaussian noise. Later, randomly varying coupling and communication constraints in complex networks were handled by an asynchronous estimator in \cite{Yong2017asy}, with gains derived utilizing the LMIs method. Note that the approaches in \cite{Ding2012H,shen2013H,wang2016anevent,Yong2017asy} are suitable for time-invariant systems (or those with time-invariant estimation error dynamics). Recently, to address the state estimation problem for time-varying complex dynamical networks, recursive LMIs approaches were designed on a finite horizon in \cite{dong2018variance}, where several LMIs are needed to be solved to obtain the estimator gains at every step. In general, although the  LMIs-based approaches can handle the $H_{\infty}$ state estimation problem for complex networks, they are limited by the following drawbacks: 1) the LMIs are usually solved in an augmented form, which pays a high computational price; 2) the LMIs have to be solved at every step for a time-varying system, which limits the online operations of the estimation algorithms.

The variance-constrained approach is another efficient way to handle the robust state estimation problem for complex networks \cite{hu2013recursive,HU2016A,hu2019variance,li2017nonaugmented,li2017state,li2019resilient}. In their pioneering work, Hu et al. \cite{hu2013recursive} designed an augmented estimator for a class of time-varying coupled complex networks with fading measurements by employing the Riccati-like difference equations technique. In this work, the estimation error covariance is upper bounded by a sequence of recursive matrices.  Later, this scheme was extended to cases with  time-varying complex networks in \cite{HU2016A} and  uncertain inner coupling in \cite{hu2019variance}, respectively. Note that the estimator gain matrices in \cite{HU2016A,hu2019variance} are determined in the compact form, which are solved by utilizing the global information. However, in terms of computational complexity and operation efficiency, it is preferable to resolve the gains in a distributed way for each node  \cite{li2017nonaugmented,li2017state,li2019resilient}.  For this purpose, Li at al. \cite{li2017nonaugmented} designed a non-augmented estimator by solving two Riccati-like difference equations, without the need of calculating cross-covariance matrices between coupling nodes. In particular, two inequality conditions on the Riccati-like equations had to be satisfied at every step, which limits the estimator's  applicability.  Recently, a boundedness analysis of the estimation error dynamics was presented in \cite{li2019resilient}, where a sufficient condition concerning system matrices was introduced to ensure the estimator feasibility.  However, this condition was ineffective since it contains several parameters implicitly in the estimators to be carefully designed.
Therefore, the estimators designed by variance-constrained methods are subject to strict feasibility criteria. Besides, the estimators in \cite{HU2016A,hu2019variance,li2017nonaugmented,li2017state,li2019resilient}
failed to handle the case of system parameters with uncertainties, which usually exist in physical systems, such as manipulators and artificial satellites.  As a remedy, a penalization approach was introduced to handle the parametric errors for the state estimation of networked systems in \cite{Zhou2013}. However, it was assumed that every system matrix was first-order differentiable with respect to each parametric error, and a central unit was needed to derive all estimator gains. In this sense, the algorithm in \cite{Zhou2013} is not fully distributed.

This paper focuses on the state estimation problem for complex coupled networks with deterministic matrix uncertainties, Gaussian noise and nonlinearities in each node's dynamics. Following the idea of regularized least-squares for a time-invariant single system   \cite{SAYED2002A} and using a decoupling technique, the problem  is reformulated as an optimization problem such that the disturbances and nonlinearities can be simultaneously modeled under a unified framework. Then, based on the measurements of each node, a novel class of estimators is presented with gains solved in a distributed way. Moreover, a boundedness analysis of the estimator parameters and mean-square  errors (MSE) is presented. Compared to \cite{SAYED2002A}, this paper faces two main technical challenges: 1) the introduction of system nonlinearity will greatly increase the difficulty of the robust state estimation problem since the parametric errors and the linearization errors are heterogeneous; 2) the coupling in the networked system makes the derivation of the estimator gains more difficult, where the information of neighbors needs to be evaluated accurately and utilized effectively. Moreover, the analysis of the estimation performance, together with the feasibility of the estimator, is particularly challenging.

The main contributions of this paper are three-fold. \textit{First}, the proposed estimators  can deal with  time-varying nonlinear system dynamics with norm-bounded system matrix uncertainties, more general than those in the literature. Second, by using a decoupling technique, the estimator gains are solved for each node individually, i.e., in a distributed way rather than in an augmented form as the LMIs-based approaches \cite{Ding2012H,shen2013H,wang2016anevent,Yong2017asy,Wu2018state,shen2011bounded,dong2018variance}. Thus, the proposed estimators have the superiority in terms of the computational complexity. Third, based on a rigorous boundedness analysis of MSE, one criterion is derived to ensure the feasibility of the estimators, which is easier to check than those from variance-constrained approaches \cite{HU2016A,hu2019variance,li2017nonaugmented,li2017state}.

The rest of this paper is organized as follows. In section \ref{sec2}, some notations, useful lemmas and the problem statement are presented. In section \ref{sec3}, a class of robust state estimators for complex networks are designed in a distributed fashion, respectively. Both the feasibility of the estimators and the estimation performance are analyzed. In section \ref{sec4}, a numerical simulation is provided to verify the effectiveness of the theoretical results. In section \ref{sec5}, some conclusions are drawn.

\section{Preliminaries and Problem Statement}\label{sec2}

\subsection{Notations and Useful Lemmas}\label{sec2.1}
\textit{Notations}: $P^T$, $\|P\|_2$ and $P^{-1}$ stand for the transpose, the 2-norm and the inverse of matrix $P$, respectively. $P>Q$ means that $P-Q$ is a positive definite matrix.
For a matrix $P \in \mathbb{R}^{n \times n}$ and a vector $q \in \mathbb{R}^{n \times 1}$, $\|q\|_{P}^2 $ denotes $q^T  P q $. $\text{diag} \{ \cdot \}$ represents a block-diagonal matrix. $\mathbb{E} \{ \cdot \}$ is the mathematical expectation of a random variable.
$\text{Tr}(\cdot)$ denotes the trace a matrix.
$I_n$ is the $n \times n$ identity matrix.
$0_n$ is the $n \times n$ zero matrix.
$\sigma_{max} \{ \cdot \}$ denotes the maximum singular value of a matrix.
$S_{\lambda}^{Re} ( \cdot )$ is the set of eigenvalues' real parts of a square matrix, and $ \overline{S}_{\lambda}^{Re} ( \cdot ) $ is the set of the absolute values of elements in $S_{\lambda}^{Re} ( \cdot )$.

\vspace{6pt}

\begin{lemma} \label{lemma1} \cite{zhou1996robust}
For any given matrices $P \in \mathbb{R}^{n \times n}$, $Q\in \mathbb{R}^{m \times m}$ and $S \in \mathbb{R}^{m \times n}$, if $P^{-1}$ and $Q^{-1}$ exist, then
\begin{align}
 & (P + S^T Q S)^{-1}
 =  P^{-1} - P^{-1} S^T ( Q^{-1} + S P^{-1} S^T )^{-1} S P^{-1}.  \notag
\end{align}
\end{lemma}

\vspace{6pt}
\begin{lemma} \label{lemma2}
For any given matrices $P \in \mathbb{R}^{n \times m}$ and $Q  \in \mathbb{R}^{n \times m}$, and a positive scalar $\beta$, one has
\begin{align}
  (P+Q)^T(P+Q) \leq  (1 + \beta)P^T P + (1 + \beta^{-1})Q^T Q. \notag
\end{align}
\end{lemma}

\vspace{6pt}
\begin{lemma} \label{lemma3}
For any given matrices $P_i \in \mathbb{R}^{n \times m}$, $i =1$, $2$, $\ldots$, $N$, one has
\begin{align}
  \left( { \sum_{i=1}^{N} P_i } \right )^T \left( { \sum_{i=1}^{N} P_i } \right )  \leq  N \sum_{i=1}^{N}  P_i^T P_i. \notag
\end{align}
\end{lemma}

\subsection{Problem Statement}\label{sec2.3}
Consider a class of discrete-time uncertain nonlinear networks consisting of $N$ nodes (agents) as follows:
\begin{align}
x_{k+1,i} & = f(x_{k,i})  + a \sum_{j \in \mathcal{N}_i} \pi_{ij} g(x_{k,j}) + \omega_{k,i} , \notag  \\
y_{k,i} & = h_i(x_{k,i}) + \nu_{k,i}, \ \ i=1, \ \cdots, \ N ,  \notag
\end{align}
where $x_{k,i} \in \mathbb{R}^{n}$ is the system state of node $i$ at step $k$, $y_{k,i} \in \mathbb{R}^{r}$ is the measurement at step $k$, $a$ is a positive constant, $\pi_{ij}$ is the coupling strength between nodes with $\pi_{ij} \neq 0$ if $j \in \mathcal{N}_i$ (the set of node $i$'s neighbors, including node $i$ itself), otherwise $\pi_{ij}=0$, $\omega_{k,i} \in \mathbb{R}^{n} $ is the process white Gaussian noise with covariance $Q_{k,i}>0$ while $\nu_{k,i} \in \mathbb{R}^{r}$ is the measurement white Gaussian noise with covariance $R_{k,i}>0$, $\omega_{k,i}$ and $\nu_{k,i}$, $i=1$, $\cdots$, $N$, $k=1$, $2$,  $\cdots$, are uncorrelated, and $f(\cdot)$, $g(\cdot)$ and $h_i(\cdot)$ are nonlinear system functions.

In some physical systems, such as cooperative unmanned aerial vehicles and multi-manipulator systems, the exact processing dynamics is difficult or even impossible to know due to imprecise system parameters and structures. This likely gives rise to the inaccuracy in $f(\cdot)$, $g(\cdot)$ and $h_i(\cdot)$ with the system model errors being bounded by positive definite matrices \cite{xiong2012robust,xiao2016tracking}. Under such situations, taking $f(\cdot)$ for example, it is more feasibly described by
\begin{align}
f(x_{k,i}) & = \hat{f}(x_{k,i}) + \Delta f_{k,i} x_{k,i} ,\notag
\end{align}
where $\hat{f}(\cdot)$ is the known nominal system function,  $\Delta f_{k,i} x_{k,i}$ is the structured uncertain term with $\Delta f_{k,i} = E_{1,k,i} \Delta_{1,k,i} $, in which $E_{1,k,i}$ is a known matrix and $\|\Delta_{1,k,i}\|_2 \leq 1$, representing the structure and amplitude of the uncertainty, respectively. In the following,  $E_{1,k,i}$ plays a key role in restraining system perturbation and ensuring   estimation performance. Although such a model cannot characterize all uncertain nonlinear systems, it can describe a large number of physical plants, such as Lur'e systems and all those satisfying the Lipschitz condition.

In this paper, the developed method for dealing with the nonlinearity and uncertainty of $f(\cdot)$ can be directly extended for $g(\cdot)$ and $h_i(\cdot)$.  Hence, for simplicity, $g(x_{k,i})$ and $h_i(x_{k,i})$ are assumed to be $G x_{k,i}$  and  $H_{i} x_{k,i}$, respectively, with known matrices $G$ and $H_{i}$. On the basis of the above discussion, the system dynamics is modified to be
\begin{align} \label{equ:observed_system}
x_{k+1,i} & = \hat{f}(x_{k,i}) + \Delta f_{k,i} x_{k,i}  + a \sum_{j \in \mathcal{N}_i} \pi_{ij} G  x_{k,j} + \omega_{k,i}  , \\
\label{equ:sensors} y_{k,i} & = H_{i} x_{k,i} + \nu_{k,i}, \ \ i=1, \ \cdots, \ N.
\end{align}

\vspace{6pt}

To deal with the state estimation problem for the above nonlinear networks and measurements containing noise, uncertainty, and nonlinearity, some   methods were developed  \cite{theodor1996robust,li2017nonaugmented,HU2016A}, which are summarized as follows.

First, denote the priori estimate and the posteriori estimate by  $\overline{x}_{k,i}$ and $\hat{x}_{k,i}$, respectively.
Here, by utilizing previous estimates and current measurements, the structure for estimates at the current step is designed as
\begin{align} \label{equ:overline_x}
\overline{x}_{k+1,i} & = \hat{f}(\hat{x}_{k,i}) + a \sum_{j \in \mathcal{N}_i} \pi_{ij} G  \hat{x}_{k,j}  ,  \\
 \label{equ:hat_x} \hat{x}_{k+1,i} & = \overline{x}_{k+1,i} + K_{k+1,i} (y_{k+1,i} - H_{i} \overline{x}_{k+1,i}),
\end{align}
where $K_{k+1,i}$ is the estimator gains to be optimized.  This structure is proposed for good reasons. First, the priori estimate is designed by following the same structure as that of the plant. Then, this value is modified to obtain the posteriori estimate by using the innovation error of node $i$. If innovation errors of neighbors are adopted to develop the estimator, then neighbors' information of every neighbor is needed to derive the estimator gains of node $i$, which consumes much more communication resources.  Besides, the introduction of such global information leads to complex coupled terms, making it practically impossible to solve the estimator gains in a distributed way.

Then, define the estimator error at step $k$ as $e_{k,i} = x_{k,i} - \hat{x}_{k,i}$. Note that the normal nonlinear function satisfies
\begin{align}  \label{equ:nonlinear}
\hat{f}(x_{k,i}) - \hat{f}(\hat{x}_{k,i}) = F_{k,i} e_{k,i} + O(e_{k,i}),
\end{align}
where $ F_{k,i} = \frac{\partial \hat{f}(x)}{ \partial x}|_{\hat{x}_{k,i}} $ and $O(e_{k,i})$ represents the high-order terms.  In many physical systems, $O(e_{k,i})$ can be described by \cite{HU2016A,li2017state,zhou1996robust}
\begin{align} \label{equ:nonlinearuncertain}
O(e_{k,i}) =  E_{2,k,i} \Delta_{2,k,i} e_{k,i},
\end{align}
where $E_{2,k,i}$ is a known matrix and $\Delta_{2,k,i}$ is the uncertain matrix that satisfies $ \| \Delta_{2,k,i} \|_2 \leq 1 $.

Based on the above process for dealing with nonlinearities, the dynamics of the estimation error at step $k+1$ can be derived as
\begin{align}
e_{k+1,i}  = & \Big{[} I - K_{k+1,i} H_{i} \Big{]} \Big{[} \Delta f_{k,i} x_{k,i} + (F_{k,i}  + E_{2,k,i} \Delta_{2,k,i} ) e_{k,i} \notag \\
& + \omega_{k,i}   + a \sum_{j \in \mathcal{N}_i} \pi_{ij} G  e_{k,j}  \Big{]} - K_{k+1,i}  \nu_{k+1,i} . \notag
\end{align}

Now, define MSE at step $k+1$ for node $i$ as $M(e_{k+1,i})= \mathbb{E} \{  e_{k+1,i}^T e_{k+1,i} \}$ and an auxiliary matrix  as $P(e_{k+1,i}) = \mathbb{E} \{ e_{k+1,i} e_{k+1,i}^T \}$. It is worth noting that $M(e_{k+1,i}) = \text{Tr} ( P(e_{k+1,i}))$.  Then, the estimator gains $K_{k+1,i}$, $i=1, \ \cdots, \ N$, are optimized by
\begin{align}
\{ K_{k+1,i} \}_{optimal}  = \arg \min_{K_{k+1,i}} \text{Tr} ( P(e_{k+1,i})),  \quad  \quad \forall i \in \mathcal{N}, \notag
\end{align}
Since $P(e_{k+1,i})$ contains the error-estimate coupling term $e_{k,i} \hat{x}_{k,i}^T$, the linearization error $ E_{2,k,i} \Delta_{2,k,i} e_{k,i}$ and the uncertain term $\Delta f_{k,i} x_{k,i}$,  some matrix inequality techniques have to be employed. For example, in \cite{li2017nonaugmented,li2017state}, Young's inequality and a quadratic matrix inequality are introduced to provide a deterministic upper bound of MSE, which is generally too conservative. Thus, the estimation performance is degraded. Moreover, in those methods, some complex conditions have to be checked to ensure the estimator feasibility at every estimation step.
In \cite{shen2013H,dong2018variance}, $P(e_{k+1,i})$, $i=1, \ \cdots, \ N$, are developed in compact forms, and the estimation gains $K_{k+1,i}$, $i=1, \ \cdots, \ N$, are optimized by solving several higher-order LMIs. In this sense, these algorithms are centralized and the computational complexity is relatively high.

To address the above limitations, this paper focuses on the following two problems.

\vspace{6pt}

\textbf{Problem 1}: How to design a distributed estimator of low computational complexity in the simultaneous presence of noise, system  uncertainties and nonlinearities?

\vspace{6pt}

\textbf{Problem 2}: How to guarantee the feasibility of the proposed estimator and the estimation performance?

\section{Main results}\label{sec3}
In this section, a novel state estimation framework is introduced for dynamic networks (\ref{equ:observed_system}) subject to Gaussian noise, nonlinearities and uncertainties. The boundedness of the estimator gain is analyzed. The estimation performance is theoretically evaluated.

\subsection{Design of Distributed State Estimators}\label{sec3.2}
In this subsection, inspired by the regularized least-squares problem with uncertainties discussed in \cite{SAYED2002A}, the state estimation problem for the uncertain nonlinear networks (\ref{equ:observed_system}) is investigated in
the following distributed form.

\vspace{6pt}

First, introduce an approximate covariance matrix for node $i$ at step $k+1$ as
\begin{align} \label{equ:P2}
P_{k+1,i}  = & (1+a)( \breve{P}_{k+1,i}^{-1}  + H_{i}^{T} \hat{R}_{k+1,i}^{-1} H_{i} + 2 \overline{\lambda}_{k,i} I_n)^{-1} +  (a+a^2) \notag \\
 \times (N  - & 1)   \sum_{j \in \mathcal{N}_i, j \neq i}  \pi_{ij}^2  G  ( P_{k,j}^{-1} +    G^T  H_{i}^T Z_{k,ij} H_{i} G)^{-1} G^T,
 \end{align}
with
\begin{align}
 \label{equ:breve_P2} \breve{P}_{k+1,i} = & (F_{k,i} + a \pi_{ii} G) P_{k,i} (F_{k,i} + a \pi_{ii} G)^T + Q_{k,i},\\
\hat{R}_{k+1,i}  = & (1+a)^{-1} R_{k+1,i} - \overline{\lambda}_{k,i}^{-1} H_{i}  E_{k,i} E_{k,i}^{T}  H_{i}^{T},  \notag  \\
Z_{k,ij} = & \pi_{ij}^2(a+a^2)  (N-1) R_{k+1,i}^{-1}, \notag
\end{align}
where  $E_{k,i} =[E_{1,k,i}, \ E_{2,k,i}] $, parameters $\overline{\lambda}_{k,i}$ and $\alpha_{k,i} \in [0$, $1]$ are two scalars to be further designed in the last part of this subsection. In this paper, $P_{k,i}$  and $\breve{P}_{k,i}$ are not real estimation posterior and priori error covariance matrices, respectively. Instead, they are iterated  similarly to those of the standard Kalman filter, in order to act as the real covariance matrices to evaluate the posterior and priori  error statistical characteristics. In the following subsections, it will be proved that they indeed guarantee the boundedness of the estimator gains and the estimation performance.

Then, design  the following distributed estimator.
\begin{align} \label{equ:estimate3}
  & \hat{x}_{k+1,i} =  (1 - \alpha_{k,i} )[\overline{x}_{k+1,i}^{u} +  K_{1,k+1,i}  b_{k+1,i}^{u}]  + {\alpha}_{k,i} \overline{x}_{k+1,i} \notag \\
  & \qquad + \sum_{j \in \mathcal{N}_i, j \neq i} \bigg [ {\alpha}_{k,i} K_{2,k+1,ij}  b_{k+1,i} \Big /  \sum_{j \in \mathcal{N}_i, j \neq i} \pi_{ij}\bigg ],
\end{align}
where
\begin{align} \label{equ:overline_xd}
  \overline{x}_{k+1,i}  = &  \hat{f}(\hat{x}_{k,i}) + a \sum_{j \in \mathcal{N}_i} \pi_{ij} G  \hat{x}_{k,j}, \notag  \\
  \overline{x}_{k+1,i}^{u}  = &  \overline{x}_{k+1,i}  -   \overline{\lambda}_{k,i}  (F_{k,i} + a \pi_{ii} G)  P_{k,i} \hat{x}_{k,i} /(1 - \alpha_{k,i} ) ,  \\
  b_{k+1,i} = &  y_{k+1,i} - H_{i}  \overline{x}_{k+1,i} , \notag \\
  b_{k+1,i}^{u} = &  y_{k+1,i} - H_{i}  \overline{x}_{k+1,i}^{u}  ,\notag \\
  K_{1,k+1,i}  & =  ( \breve{P}_{k+1,i}^{-1}  + H_{i}^{T} \hat{R}_{k+1,i}^{-1} H_{i} + 2 \overline{\lambda}_{k,i} I_n  )^{-1}   H_{i}^{T}  \hat{R}_{k+1,i}^{-1} ,\notag \\
   K_{2,k+1,ij} &  =   a \pi_{ij}  G ( P_{k,j}^{-1} +   G^T  H_{i}^T Z_{k,ij} H_{i} G)^{-1}   G^T H_{i}^T Z_{k,ij}. \notag
\end{align}

\vspace{6pt}

With the appropriate parameters  $\overline{\lambda}_{k,i}$ and $\alpha_{k,i}$, the proposed state estimation algorithm for the  uncertain nonlinear networks in a distributed sense is summarized as Algorithm \ref{algorithm2}. The derivation of Algorithm \ref{algorithm2} is given in Appendix \ref{proof1}.

\begin{algorithm}[t]
\caption{State estimation for node $i$ at step $k+1$ by the  distributed method}
\hspace*{0.02in} {\bf Initialization:} The parameters $\overline{\lambda}_{0,i}$ $\breve{P}_{0,i}$ and $P_{0,i}$ are set as:
\begin{align}
\overline{\lambda}_{0,i} & = (1 + \beta) \| E_{0,i}^{T}  H_{0,i}^{T} R_{0,i}^{-1} H_{0,i} E_{0,i}\|_2,  \ \beta > 0, \notag \\
0 < &\breve{P}_{0,i} \leq Q_{0},  \notag \\
P_{0,i} & = ( \breve{P}_{0,i}^{-1} + H_{0,i}^T \hat{R}_{0,i}^{-1} H_{0,i})^{-1}. \notag
\end{align}

\hspace*{0.02in} {\bf Input:} $y_{k+1,i}$, \  $\hat{x}_{k,j}$ and $P_{k,j}$, $j \in \mathcal{N}_i $;

\hspace*{0.02in} {\bf Output:} $\hat{x}_{k+1,i}$ and $P_{k+1,i}$;

\label{algorithm2}
\begin{algorithmic}[1]
\State Exchange the information package $(\hat{x}_{k,i}$, $P_{k,i})$ with its neighbors;

\State Calculate the recursive matrix $P_{k+1,i}$ by (\ref{equ:P2});

\State Get the priori state estimates $\overline{x}_{k+1,i}$ and $ \overline{x}_{k+1,i}^{u}$ by (\ref{equ:overline_xd});

\State Compute the posterior state estimate $\hat{x}_{k+1,i}$ by (\ref{equ:estimate3});

\end{algorithmic}
\end{algorithm}

\vspace{6pt}

Actually, Algorithm \ref{algorithm2} is designed based on the following optimization problem.
\begin{align}   \label{equ:optimal_uncertain1}
 & \qquad \qquad \min_{\stackrel{x_{k,j} } {w_{k,i}  }} J_{1,i}(x_{k,j}, \ w_{k,i}, \ x_{k+1,i}, \ j \in \mathcal{N}_i)  , \\
  &  \text{s.t.} \quad x_{k+1,i}  = \hat{f}(x_{k,i}) + \Delta f_{k,i} x_{k,i}  + a \sum_{j \in \mathcal{N}_i} \pi_{ij} G  x_{k,j} + w_{k,i} , \notag
\end{align}
with
\begin{align}   \label{equ:J11}
 J_{1,i} = &    \| x_{k,i} - \hat{x}_{k,i} \|^2_{P_{k,i}^{-1}}  + \| w_{k,i} \|^2_{Q_{k,i}^{-1}} +  \| y_{k+1,i} - H_{i} x_{k+1,i} \|^2_{R_{k+1,i}^{-1}} \notag \\
  &   +   \sum_{j \in \mathcal{N}_i} \| x_{k,j} - \hat{x}_{k,j} \|^2_{P_{k,j}^{-1}},
\end{align}
where $P_{k,i}$ is designed in (\ref{equ:P2}), and $Q_{k,i}$ and $R_{k+1,i}$ are given in Section \ref{sec2.3}. It is worth mentioning that $x_{k,j}$ and $w_{k,i}$ in (\ref{equ:optimal_uncertain1}) are optimization variables of the problem, instead of the real state and the process noise.  The structure of $ J_{1,i}$ in (\ref{equ:J11}) can be interpreted as follows. The first three parts of $J_{1,i}$ are optimized to obtain the priori estimate according to the priori information, such as the system model and the noise covariance. Then, by adding the last term, the local measurement information is utilized efficiently to correct the priori estimate. Hence, the cost function attempts to balance the model prediction process and the measurement feedback process.

\vspace{6pt}
\begin{remark}
The parameter $\alpha_{k,i}$ is introduced to act as a regulatory indicator, balancing the weights of the innovation errors on node $i$ and its neighbors. Specifically, the priori estimate of node $i$ is corrected by $(1 - \alpha_{k,i}) b_{k+1,i}$ while the ones of node $i$'s neighbors are corrected by $\alpha_{k,i} b_{k+1,i} /(\sum_{j \in \mathcal{N}_i, j \neq i} \pi_{ij}) $. Compared with the existing methods putting node $i$'s own and neighbors' information together, this decoupling design simplifies the solution process. Meanwhile, by this technique, the impact of the whole network on the subsystems is decoupled onto the neighbors at each estimation step. Here, how to obtain the optimized $\alpha_{k,i}$ is the key to guarantee the decoupling effect, which will be studied in the following part.
\end{remark}

\vspace{6pt}

\begin{remark}
Compared with the estimation algorithm in \cite{Zhou2013}, Algorithm \ref{algorithm2} possesses the following   differences and advantages. First, the estimator gain of each node is developed  based only on neighbors' information, different from that in \cite{Zhou2013}, where a central unit is needed to collect global information. Second, every system matrix is required to be first-order differentiable with respect to each parametric error in \cite{Zhou2013}, while the system matrix uncertainties in this paper are described by bounded parameters based on graph theory. Third, in the presence of system uncertainties, rigorous proofs for the feasibility of the estimators and the boundedness of MSE are provided in this paper.
\end{remark}

\vspace{6pt}
\begin{proposition}
The computational complexity of Algorithm \ref{algorithm2}  is $O(|\mathcal{N}_i| n^3 + r^3)$, where $\mathcal{N}_i$ is the neighbor set of node $i$.
\end{proposition}

\vspace{6pt}

\begin{proof}
Note that Algorithm \ref{algorithm2} consists of three recursive equations (\ref{equ:P2}), (\ref{equ:estimate3}) and (\ref{equ:overline_xd}) with several auxiliary matrices, and that $P_{k,i} \in \mathbb{R}^{n \times n}$, $G \in \mathbb{R}^{n \times n}$, $ \hat{R}_{k+1,i} \in \mathbb{R}^{r \times r}$ and $H_{i} \in \mathbb{R}^{r \times n}$ in (\ref{equ:P2}). According to \cite{arora2009computational}, the computational complexity of (\ref{equ:P2}) can be evaluated as $O(|\mathcal{N}_i| n^3 + r^3)$. Similarly, the ones  of (\ref{equ:estimate3}) and (\ref{equ:overline_xd}) are $O(|\mathcal{N}_i| n^3 + r^3)$ and $O( n^3 )$, respectively.
Hence, the total computational complexity of Algorithm \ref{algorithm2}  is $O(|\mathcal{N}_i| n^3 + r^3)$.
\end{proof}

\vspace{6pt}

\begin{remark}
Compared with the centralized algorithms in \cite{HU2016A,hu2019variance}, whose computational complexity is  $O(N^3 n^3 + N^3 r^3)$, Algorithm \ref{algorithm2} in this paper shows great superiority of saving computational resources.
\end{remark}

\vspace{6pt}

In the following, an optimization approach is presented for determining parameters $\overline{\lambda}_{k,i}$ and $\alpha_{k,i}$ in (\ref{equ:estimate3}).

\vspace{6pt}

\begin{theorem}\label{thm3}
Consider networks (\ref{equ:observed_system}) with measurements (\ref{equ:sensors}). The parameters $\overline{\lambda}_{k,i}$ and $\alpha_{k,i}$ in estimator (\ref{equ:estimate3}) can be obtained by
\begin{align} \label{equ:alpha_lambda}
 \{ \alpha_{k,i}^{opt},  \ \overline{\lambda}_{k,i}^{opt} \} = &  \arg \min_{ \stackrel{\alpha_{k,i} \in [0, \ 1]} { \overline{\lambda}_{k,i} \ge \| H^{T} T H \|_2 } } \hat{J}_{2,i}( \overline{\lambda}_{k,i}, \alpha_{k,i}),
\end{align}
where $\hat{J}_{2,i}( \overline{\lambda}_{k,i}, \alpha_{k,i})$ is given in (\ref{equ:J_Alpha_L}) in the Appendix.

\end{theorem}

\vspace{6pt}
The proof of Theorem \ref{thm3} is given in Appendix \ref{proof2}.
\vspace{6pt}

\begin{remark}
Although $\hat{J}_{2,i}( \overline{\lambda}_{k,i}, \alpha_{k,i})$ in (\ref{equ:alpha_lambda}) is not convex with respect to $\overline{\lambda}_{k,i}$ and $\alpha_{k,i}$, some existing algorithms can be used to solve the problem, such as the quasi-Newton
methods \cite{BartholomewBiggs2008Nonlinear,convexoptimization}. Generally, the solving  methods for (\ref{equ:alpha_lambda}) are mature and their computational complexity is acceptable.
\end{remark}

\vspace{6pt}

As shown in \cite{sayed2001framework} and \cite{zhou2010on}, one can  set $\overline{\lambda}_{k,i}=(1+\beta) \breve{\lambda}_{k,i}$ for computational simplification, where $\beta \in (0$, $1)$, and
\begin{align} \label{equ:breve_lambda}
 & \breve{\lambda}_{k,i} =  (1+a)^{-1} \| E_{k,i}^{T}  H_{i}^{T} R_{k+1,i}^{-1}   H_{i}  E_{k,i} \|_2.
\end{align}
For example, in \cite{zhou2010on}, $\beta=0.5$ usually generates a desirable estimation performance. For linear time-invariant systems, when the estimator gains converge, $\overline{\lambda}_{k,i}$ tends to a fixed value. Thus, in such situations, $\overline{\lambda}_{k,i}$ can be pre-computed to achieve a satisfactory suboptimal estimation. Further, the analytical expression of $\alpha_{k,i}^{opt}$  is given as follows.

\vspace{6pt}

\begin{corollary} \label{cor1}
If $\overline{\lambda}_{k,i}=(1+\beta) \breve{\lambda}_{k,i}$,  then $\alpha_{k,i}^{opt} $ can be designed by
\begin{align} \label{equ:alpha_opt}
 & \alpha_{k,i}^{opt} = \begin{cases}
 \ 0 & \text{if $ \hat{\Phi}_{k,i} < 0$ },\\
  \hat{\Phi}_{k,i} & \text{if $ \hat{\Phi}_{k,i} \in [0$, $1]$ },\\
   \  1 & \text{if $ \hat{\Phi}_{k,i} > 1$ },\\
  \end{cases}
\end{align}
with
\begin{align}
 & \hat{\Phi}_{k,i} =( \Phi_{k,i}^{1} + 2  \Phi_{k,i}^{2} ) / \left ({ 2 \Phi_{k,i}^{2} + 2 \sum_{j \in \mathcal{N}_i, j \neq i} \Phi_{k,ij}^{3} }  \right ), \notag
\end{align}
where $\Phi_{k,i}^{1}$, $\Phi_{k,i}^{2}$, and $\Phi_{k,ij}^{3}$ are defined in (\ref{equ:Phi}) in the Appendix.
\end{corollary}

\vspace{6pt}
The proof of Corollary \ref{cor1} is given in Appendix \ref{proof3}.
\vspace{6pt}

From (\ref{equ:alpha_opt}), $\Phi_{k,i}^{1}$ is related to the level of system perturbations while $\Phi_{k,ij}^{3}$ is related to the confidence of node $i$'s neighbors' information. If system perturbations are larger, which leads to a larger $\Phi_{k,i}^{1}$, then $\alpha_{k,i}^{opt}$ is larger. Similarly,  if the neighbors' information is less confident, which leads to a larger $\Phi_{k,ij}^{3}$, then $\alpha_{k,i}^{opt}$ is smaller. In this sense, $\alpha_{k,i}$ acts as a regulatory indicator to  determine the weight of the residual error for each node's own and its neighbors' information. In particular, if there are no system uncertainties and neighbors, i.e., $\Delta f_{k,i}=0$ and $a=0$, then the estimation problem becomes the standard Kalman filtering problem. In such case, one can show that $\Phi_{k,i}^{1}=0$ and $\Phi_{k,ij}^{3}=0$, thus, $\alpha_{k,i}^{opt}=1$, which is consistent with the standard Kalman filter.

\vspace{6pt}

\begin{remark}
In this paper, the homogeneity  of $f(\cdot)$ and $g(\cdot)$  for each node is not needed for the derivation of Algorithm \ref{algorithm2}. Instead, it is required to guarantee the estimation performance by Algorithm \ref{algorithm2} under a much easier-to-check condition compared to the literature \cite{li2017nonaugmented,HU2016A,li2017state}, which will be clarified in the following subsections. Even though the two functions are same for all nodes, their Jacobian matrices, e.g.,  $F_{k,i} = \frac{\partial \hat{f}(x)}{ \partial x}|_{\hat{x}_{k,i}} $, $ i=1,  \cdots, N$, are not identical due to different $\hat{x}_{k,i}$.  In this sense, the model considered in this paper is heterogeneous, which is as general as the time-varying linear models adopted in \cite{Zhou2013,hu2013recursive} and the nonlinear models adopted in \cite{li2017nonaugmented,li2017state}.
\end{remark}

\subsection{Boundedness Analysis of Estimator Gains}\label{sec3.3}
In this subsection, the boundedness of the distributed estimator gains in (\ref{equ:estimate3}) is studied. The key issue is to analyze the uniform boundedness of $P_{k,i}$, $\breve{P}_{k,i}$ and $\hat{P}_{k,i}$. Without losing generality, it suffices to study the one of $P_{k,i}$ for any $k$.

\vspace{6pt}

\begin{assumption} \label{asm1}
The matrix $F_{k,i} + a \pi_{ii} G$ is non-singular.
\end{assumption}

\vspace{6pt}

\begin{remark}
This assumption is automatically satisfied if the model and its Jacobian matrix $F_{k,i} + a \pi_{ii} G$ are obtained by the discretization of a continuous-time system \cite{BATTISTELLI201875}. Besides,   this assumption can be relaxed to be that the matrix $F_{k,i}^{0}$ is non-singular, where $F_{k,i}^{0} $ is a matrix in the ``neighbourhood" of $F_{k,i} + a \pi_{ii} G$, i.e., $\| F_{k,i} + a \pi_{ii} G - F_{k,i}^{0}  \|_2 \leq \epsilon_1 $ with $\epsilon_1 $ being a small positive number. In Assumption \ref{asm1}, $F_{k,i} + a \pi_{ii} G$ instead of $F_{k,i}^{0}$ is chosen just for readability and simplicity.
\end{remark}

\vspace{6pt}

\begin{assumption} \label{asm2}
$(F_{k,i} + a \pi_{ii} G$, $ H_{i} ) $ is uniformly observable, i.e., there exists a positive scalar $\kappa$ and a positive integer  $\overline{N}$ such that
\begin{align}
 \kappa I \leq \sum_{h=0}^{\overline{N}-1}  \Psi_{k+h-1,k}^{T} H_{i}^T R_{k+1,i}^{-1} H_{i} \Psi_{k+h-1,k}   ,   \notag
\end{align}
where
\begin{align}
& \Psi_{k+h-1,k}=
  \begin{cases}
     (F_{k+h-1,i} + a \pi_{ii} G)  \cdots (F_{k,i} + a \pi_{ii} G)     & \text{if $ h \ge 1$ }, \\
   I & \text{if $ h = 0$ }. \\
  \end{cases} \notag
\end{align}
\end{assumption}
\vspace{6pt}

For linear time-invariant systems, Assumption \ref{asm2} is equivalent to that $(H $, $ F_{i} + a \pi_{ii} G) $ is observable, where $H=[H_{1}^T$, $\ldots$, $H_{N}^T]^T $.

\vspace{6pt}
\begin{assumption} \label{asm3}  \cite{li2017state,Bonnabel2015}
The parameters in (\ref{equ:observed_system}) are uniformly bounded, i.e., there exist positive constants $\kappa_{F,1}$, $\kappa_{F,2}$, $\kappa_{E,1}$, $\kappa_{E,2}$, $\kappa_H$, $\kappa_{\omega,1}$, $\kappa_{\omega,2}$, $\kappa_{\nu,1}$ and $\kappa_{\nu,2}$ such that $ \kappa_{F,1} \leq  \|F_{k,i}  + a \pi_{ii} G\|_2 \leq \kappa_{F,2}$, $\|G\|_2 \leq \kappa_{G}$, $\|E_{1,k,i}\|_2 \leq \kappa_{E,1}$, $\|E_{2,k,i}\|_2 \leq \kappa_{E,2}$, $\|H_{i}\|_2 \leq \kappa_H$, $\kappa_{\omega,1} \leq  \|Q_{k}\|_2 \leq \kappa_{\omega,2}$, and $\kappa_{\nu,1} \leq \|R_{k,i}\|_2 \leq \kappa_{\nu,2}$ for $\forall k=1$, $2$, $\ldots$, and $\forall i=1$, $\ldots$, $N$.
\end{assumption}

\vspace{6pt}

\begin{theorem}\label{thm4}
Consider  networks (\ref{equ:observed_system}) with measurements (\ref{equ:sensors}).  If Assumptions \ref{asm1}-\ref{asm3} hold and $\overline{\lambda}_{k,i}=(1+\beta) \breve{\lambda}_{k,i}$, the recursive matrix $P_{k,i}$ is uniformly upper and lower bounded, provided that
\begin{align}
\max \{ \overline{S}_{\lambda}^{Re} ( G ) \} < (a+a^2)^{-\frac{1}{2}}  (N-1)^{-1} \pi_{m}^{-1}, \notag
\end{align}
where $\pi_{m}=\max \{ \pi_{ij}, \ \forall i,j \in [1,\ \ldots,\ N] \}$.
\end{theorem}

\vspace{6pt}
The proof of Theorem \ref{thm4} is given in Appendix \ref{proof4}.
\vspace{6pt}

\begin{remark}
Theorem \ref{thm4} provides a sufficient condition for the stability of the proposed estimation algorithm. Note that, in \cite{li2017state,li2017nonaugmented}, the corresponding condition can be summarized as that one bound of MSE must be smaller than a given (or chosen) positive number at each step. Instead, in this paper, one only needs to check a condition at the beginning of the estimation process, which shows its superiority in terms of online solving the state estimation problem for online applications.
\end{remark}

\vspace{6pt}

\begin{assumption} \label{asm4}
The system (\ref{equ:observed_system}) reduces to a class of uncertain linear time-invariant complex network, i.e., $F_{k,i} = F_{i}$, $E_{1,k,i} = E_{1,i}$, $E_{2,k,i} = E_{2,i}$,  $Q_{k,i} = Q_{i}$, and $R_{k,i} = R_{i}$.
\end{assumption}

\vspace{6pt}

Generally, for slowly varying systems with relatively fast-sampling sensors, this assumption naturally holds.

\vspace{6pt}
\begin{corollary}\label{cor2}
Consider networks (\ref{equ:observed_system}) with measurements (\ref{equ:sensors}).  If Assumptions \ref{asm1}-\ref{asm4} hold, $\alpha_{k,i} = \alpha_{i}$, and $ \overline{\lambda}_{k,i} = (1+\beta) \breve{\lambda}_{k,i}$, then the recursive matrix $P_{k,i}$ is convergent.
\end{corollary}

\vspace{6pt}
The proof of Corollary \ref{cor2} is given in Appendix \ref{proof5}.

\vspace{6pt}
Built on Theorem \ref{thm4}, Corollary \ref{cor2} reveals that the convergence of the estimator gains in (\ref{equ:estimate3}) is ensured for time-invariant systems.

\subsection{Boundedness Analysis of MSE}\label{sec3.4}
In this subsection, the boundedness of MSE by the proposed distributed estimator is analyzed for two cases: 1) $\Delta f_{k,i}=0$ ; 2) $\Delta f_{k,i}  \neq 0$.

%First, denote the priori and posterior estimation error covariance for node $i$ by  the distributed estimator (\ref{equ:estimate3}) by $\overline{e}_{k,i}^d = x_{k,i} - \overline{x}_{k,i}^d $ and $e_{k,i}^d = x_{k,i} - \hat{x}_{k,i}^d $, respectively. It follows from (\ref{equ:observed_system}) and (\ref{equ:estimate3}) that
%\begin{align}
% e_{k+1,i}^d  =  & [I -  K_{k,i} H_{i}] \overline{e}_{k,i}^d  -  K_{k,i} \nu_{k+1,i}  -  a \sum_{j \in \mathcal{N}_i, j \neq i}  \pi_{ij} G e_{k,ij} ^{opt} \notag
%\end{align}
%and
%\begin{align}
% \overline{e}_{k+1,i}^d   = & (F_{k,i} + a \pi_{ii} G + E_{k,i} \Delta_{k,i} [I_n,\ I_n]^T) e_{k,i}  + a \sum_{j \in \mathcal{N}_i, j \neq i} \pi_{ij} G  e_{k,j} + \omega_{k,i}  +  ( E_{k,i} \Delta_{k,i} [I_n,\ 0_n]^T +  \hat{\overline{\lambda}}_{k,i}  (F_{k,i} + a \pi_{ii} G)  \hat{P}_{k,i}) \hat{x}_{k,i} \notag
%\end{align}
%where $K_{k,i}= (1 - \alpha_{k,i} ) P_{k+1,i} H_{i}^{T} \hat{R}_{k+1,i}^{-1}$
%
%\begin{align}
%x_{k+1,i}  = & \hat{f}(\hat{x}_{k,i}) + ( F_{k,i} + E_{k,i} \Delta_{k,i} [I_n,\ I_n]^T) e_{k,i} + \omega_{k,i} +  E_{k,i} \Delta_{k,i} [I_n,\ 0_n]^T \hat{x}_{k,i}  + a \sum_{j \in \mathcal{N}_i} \pi_{ij} G ( \hat{x}_{k,j} + e_{k,j}) \notag
%\end{align}
%\vspace{6pt}
%
%By denoting an augmented system state $\tilde{x}_{k,i}=[x_{k,i}$, $e_{k,i}]$,  $\tilde{x}_{k+1,i}$ is derived as
%\begin{align}
%\tilde{x}_{k+1,i}  = &  \notag
%\end{align}
\vspace{6pt}
First, consider the case of $\Delta f_{k,i}=0$, i.e., the uncertainty in the system function is omitted. The corresponding result on the estimation performance by the distributed estimator (\ref{equ:estimate3}) is summarized as follows.

\vspace{6pt}

\begin{theorem}\label{thm5}
Consider  networks (\ref{equ:observed_system}) with $\Delta f_{k,i}=0$ and measurements  (\ref{equ:sensors}). Then, the distributed estimator (\ref{equ:estimate3}) is unbiased. Furthermore, suppose that Assumptions \ref{asm1}-\ref{asm4} hold and $\overline{\lambda}_{k,i} = (1+\beta) \breve{\lambda}_{k,i}$. Then,  MSE for node $i$ is ultimately upper bounded if
\begin{align} \label{equ:condition_bound}
&  \| \mathcal{H} F \|_2 < \Big{(} 1+\kappa_{F,1}^{-1} \kappa_{E,2} + a \pi_{m} \kappa_{F,1}^{-1} \kappa_{G} \sqrt{N (N-1)}  \Big{)}^{-1},
 \end{align}
 where $\pi_{m}=\max \{ \pi_{ij}, \ i,j \in [1,\ \ldots,\ N]  \}$ and  $\mathcal{H} F$ is given in (\ref{equ:mathcal_H}).
\end{theorem}

\vspace{6pt}
The proof of Theorem \ref{thm5} is given in Appendix \ref{proof6}.
\vspace{6pt}

\begin{remark}
Theorem \ref{thm5} provides a condition for guaranteeing the estimation performance, i.e., the boundedness of Algorithm \ref{algorithm2}. Actually, by some calculations, this condition can be further relaxed as $\kappa_{F,2} + a  \pi_{m} \kappa_{G} \leq \epsilon_2 $ with $\epsilon_2 $ being a positive scalar in terms of $N$, $\kappa_{E,2}$, etc. This equivalently requires that the observed system should evolve slowly or the sensors should have the ability of fast sampling, which is consistent with the intuition.
Compared with the recursive methods used in \cite{li2017nonaugmented,HU2016A}, the condition for the above algorithm  only needs to be justified once while the ones in \cite{li2017nonaugmented,HU2016A} contain some parameters that must be carefully chosen at every step, e.g., $\alpha_{k,i}$ and $\beta_{k,i}$ in \cite{li2017nonaugmented}.
\end{remark}

\vspace{6pt}

When it comes to the case of $\Delta f_{k,i} \neq 0$, the following assumption is needed.
\vspace{6pt}
\begin{assumption} \label{asm5}
$\Delta f_{k,i} x_{k,i}$ is uniformly upper bounded, i.e, there exists a positive scalar $\kappa_u$ such that $ \| \Delta f_{k,i} x_{k,i} \|_2 \leq \kappa_u $.
\end{assumption}

\vspace{6pt}
\begin{corollary}\label{cor3}
Consider networks (\ref{equ:observed_system}) with measurements (\ref{equ:sensors}). If the conditions of Theorem \ref{thm5} and Assumption \ref{asm5} hold, MSE for node $i$ is ultimately upper bounded.
\end{corollary}

\vspace{6pt}

The proof of Corollary \ref{cor3} can be performed by simply combining the one in Theorem \ref{thm5} above with the one in Theorem 3 in \cite{sayed2001framework}.
%\vspace{6pt}

%\begin{remark}
%The uncertain system state $\Delta f_{k,i} x_{k,i}$ is necessary in Theorem \ref{thm6}.
%\end{remark}

\section{Simulation example} \label{sec4}
In this section, the theoretical results are visualized by a comparative simulation, where a discrete time-varying complex coupled network of four nodes is considered. Similarly to the simulation model considered in \cite{HU2016A}, the system nonlinear function in (\ref{equ:observed_system}) is set as
\begin{align} \label{equ:hat_ff}
f(x_{k,i})= \left[{
\begin{array}{*{20}{c}}
  { 0.9 x_{k,i}(1) + sin(0.5 x_{k,i}(2))}  \\
  { 0.9 x_{k,i}(2) - sin(0.5x_{k,i}(1) )}  \\
  \end{array} }\right],
\end{align}
with $x_{k,i}=[x_{k,i}(1), \ x_{k,i}(2)]^T$, and the measurement matrices are given as
\begin{align}
& H_{k,1}=[0.90 \ 0.25], \ H_{k,2}=[0.95 \ 0.65], \notag \\
& H_{k,3}=[0.90 \ 0.35], \ H_{k,4}=[0.85 \ 0.20]. \notag
\end{align}
Besides, the coefficient $a=0.1$, the coupling strength $\pi_{ij}$ is chosen as
\begin{align}
\pi_{ij} =
  \begin{cases}
     -0.3,     & \text{ $ i =j$ }, \\
   \  0.1, & \text{ $ i \neq j $ }, \\
  \end{cases} \notag
\end{align}
and the inner-coupling matrix $G=0.2I_2$. The processing and measurement noise covariances for each node are given as $Q_{k,i}=0.001I_2$ and $R_{k,i}=0.01$, $ i=1$, $\ldots$, $4$, $k=1$, $2$, $\ldots$.  The initial values of the system states and estimates are chosen as $x_{0,1}=[2.0 \ -2.8]^T$, $x_{0,2}=[2.5 \ -2.5]^T$, $x_{0,3}=[2.5 \ -2.0]^T$, $x_{0,4}=[2.0 \ -2.0]^T$, $\hat{x}_{0,i}=x_{0,i}+ 0.2i \times [\delta_i, \ \delta_i]^T$, $i=1$, $\ldots$, $4$, where $\delta_i$ is zero-mean Gaussian noise with unit covariance. Other parameters are given as $\beta=0.5$,  $\breve{P}_{0,i}=0.001$, $\alpha_{k,i}=0.1$, $\forall k=1$, $2$, $\ldots$.

\begin{figure}[!htb]
\centering
\subfigure[node 1]{
{\includegraphics[scale=0.41]{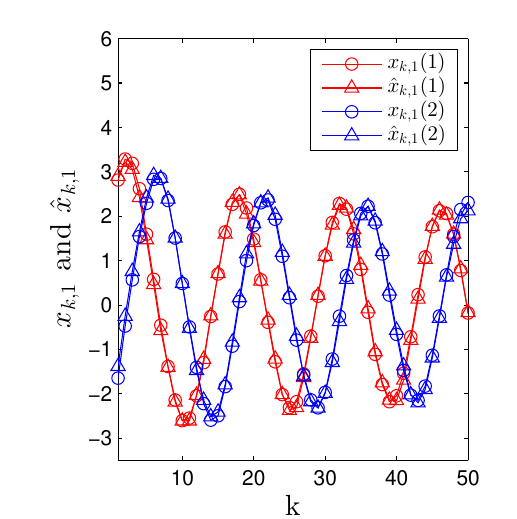}}}
\subfigure[node 2]{
{\includegraphics[scale=0.41]{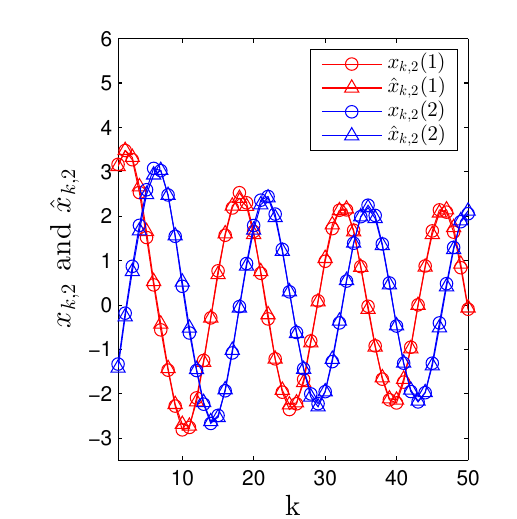}}}
\subfigure[node 3]{
{\includegraphics[scale=0.41]{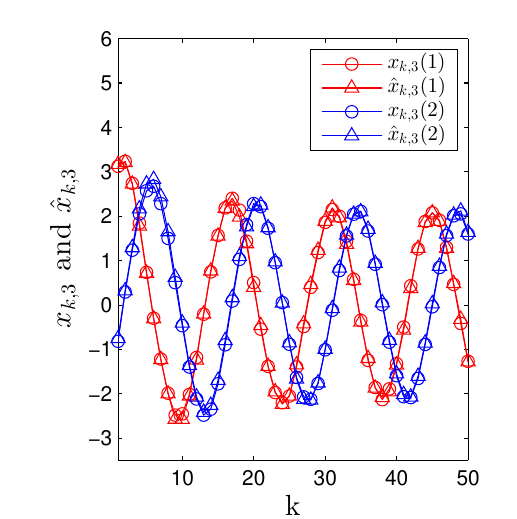}}}
\subfigure[node 4]{
{\includegraphics[scale=0.41]{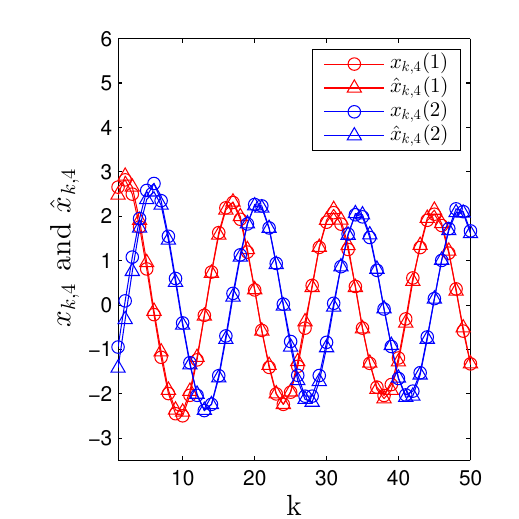}}}
\caption{The states and estimates of each node} \label{f:state}
\end{figure}

\begin{figure}[!htb]
\centering
\subfigure[node 1]{
{\includegraphics[scale=0.41]{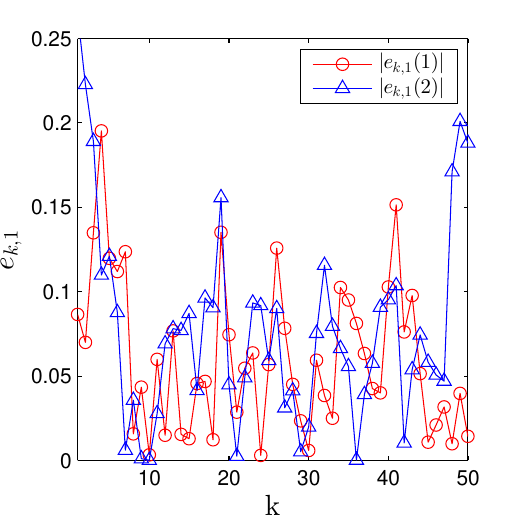}}}
\subfigure[node 2]{
{\includegraphics[scale=0.41]{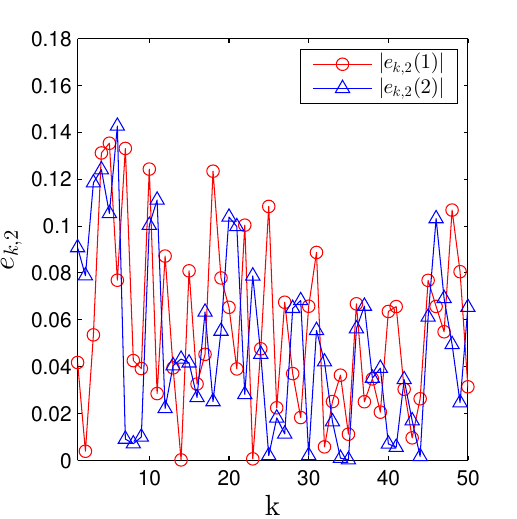}}}
\subfigure[node 3]{
{\includegraphics[scale=0.41]{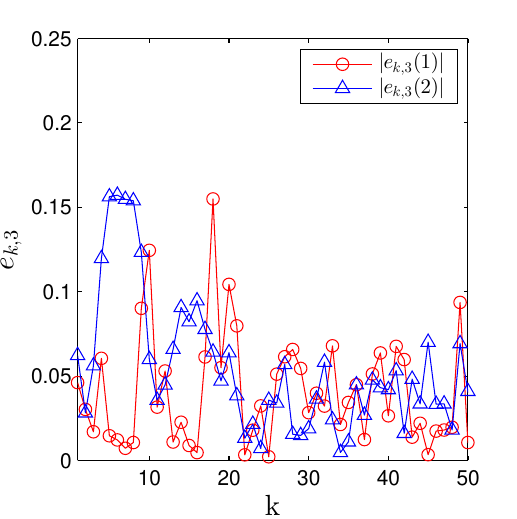}}}
\subfigure[node 4]{
{\includegraphics[scale=0.41]{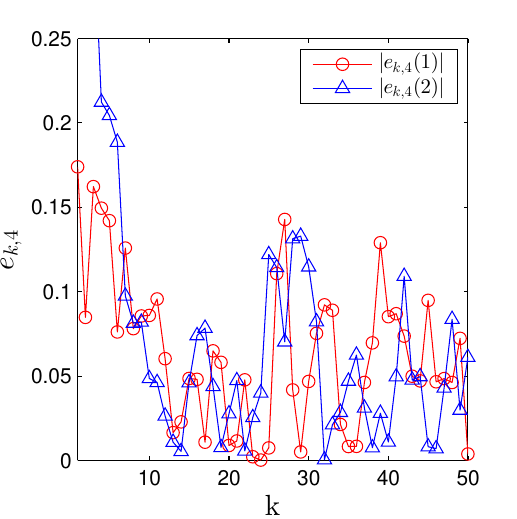}}}
\caption{The estimation errors of each node} \label{f:error}
\end{figure}

\begin{figure}[!htb]
\subfigure{{\includegraphics[scale=0.6]{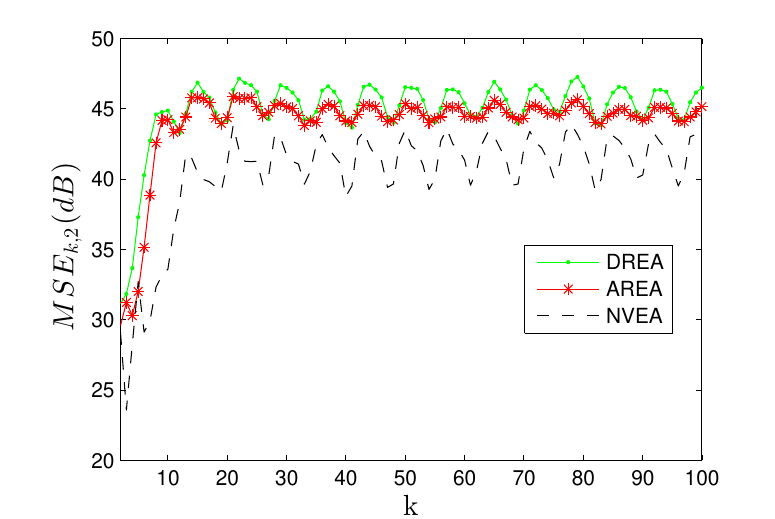}}}
\caption{The mean-square errors of node 2 without uncertainties} \label{f:com_no}
\end{figure}

\begin{figure}[!htb]
\subfigure{{\includegraphics[scale=0.6]{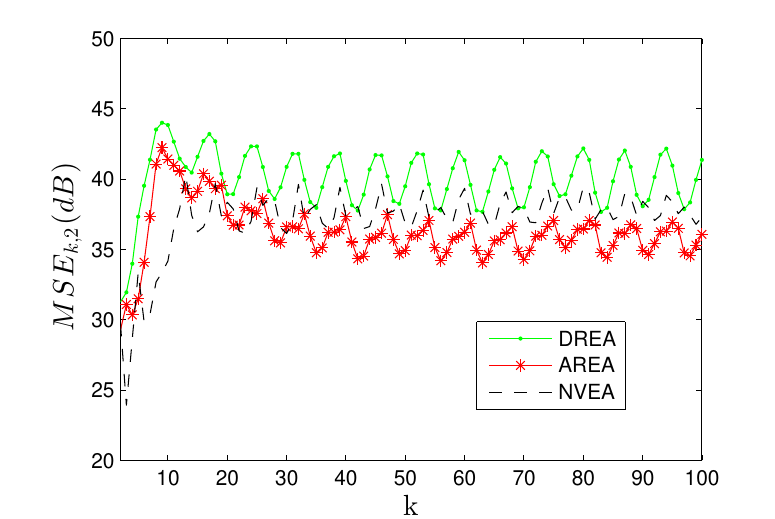}}}
\caption{The mean-square errors of node 2 with uncertainties} \label{f:com_with}
\end{figure}

\begin{figure}[!htb]
\centering
\subfigure[star]{
{\includegraphics[scale=0.34]{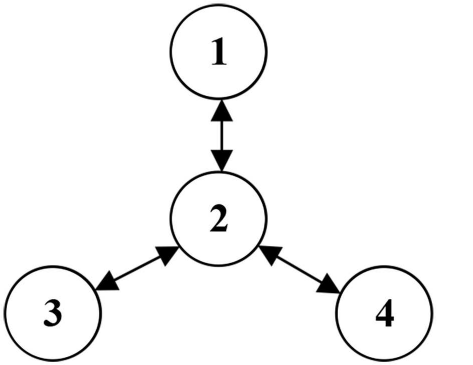}}}
\subfigure[ring]{
{\includegraphics[scale=0.33]{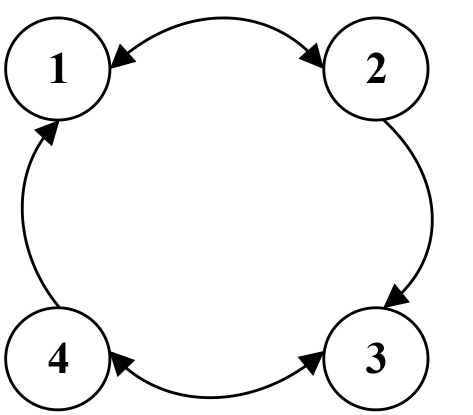}}}
\subfigure[chain]{
{\includegraphics[scale=0.40]{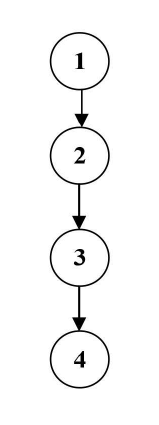}}}
\caption{ Different network topologies} \label{f:comgraph}
\end{figure}

\begin{figure}[!htb]
\subfigure{{\includegraphics[scale=0.5]{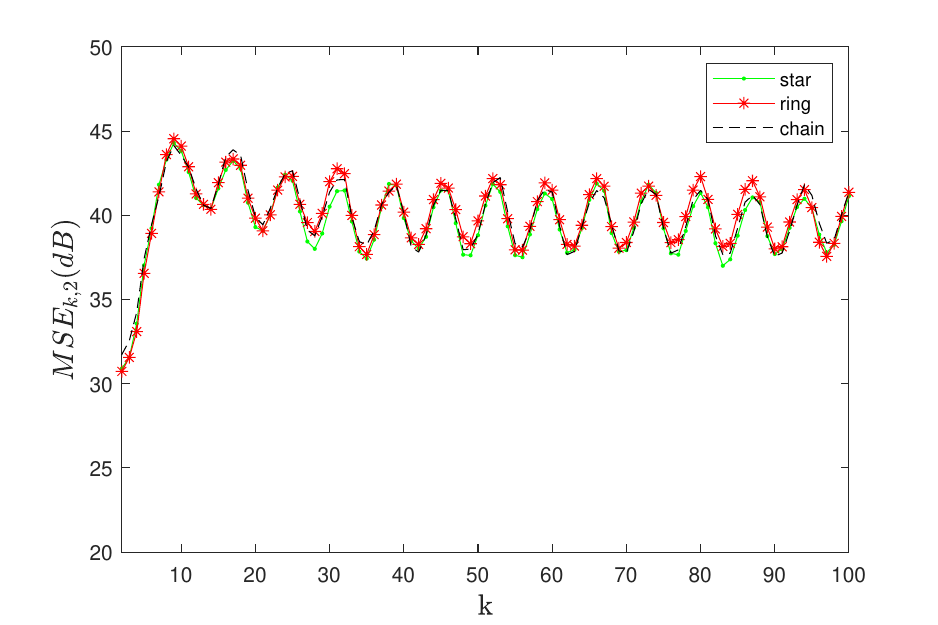}}}
\caption{ The mean-square errors of node 2 with different topologies} \label{f:graphcom}
\end{figure}

The simulation results are illustrated from two aspects: 1) the estimation performance of each node by Algorithm \ref{algorithm2}; 2) a comparative study with several typical estimation algorithms; 3) a comparative study with different network structures.

\textbf{Part 1:} By using Algorithm \ref{algorithm2} based on the above parameters, the numerical simulation results are shown in Figs.~\ref{f:state} and~\ref{f:error}, where the states, the estimates, and the estimation errors of each node are plotted. It can be seen that the estimates well follow the real states of each node in the presence of disturbances and nonlinearities, which indicates that the estimation performance is ensured by the proposed distributed estimation algorithm.

\textbf{Part 2:} To better illustrate the performance of the proposed distributed robust estimation algorithm (DREA), a comparative simulation for network (\ref{equ:observed_system}) with several typical estimation algorithms is presented, including the augmented recursive estimation algorithm (AREA) in \cite{HU2016A,hu2019variance}, the non-augmented variance-constrained estimation algorithm (NVEA) in \cite{li2017nonaugmented,li2017state,li2019resilient}. Before proceeding to the comparative results, define the average mean-square errors for node $i$ as follows.
\begin{align}
 MSE_{k,i}
 = &  \frac{1}{1000} \sum_{l=1}^{1000} \|\hat{x}_{k,i}^{l} - x_{k,i}^{l} \|_2^2,
  \notag
\end{align}
where $l$ denotes the $l$-th trial.
First, consider the dynamical network without uncertainties, i.e., $\Delta f_{k,i}=0$ in (\ref{equ:observed_system}). Without the loss of generality, the average mean-square errors (dB: $-10 \times \lg(MSE_{k,i}$)) for node $2$ by the above algorithms are shown in Fig.~\ref{f:com_no}. Then, consider the case with uncertainties, where the unknown real system nonlinear function is
\begin{align}
f(x_{k,i})= \left[{
\begin{array}{*{20}{c}}
  { 0.9 x_{k,i}(1) + sin(0.525 x_{k,i}(2))}  \\
  { 0.9 x_{k,i}(2) - sin(0.525 x_{k,i}(1) )}  \\
  \end{array} }\right], \notag
\end{align}
and the known nominal system nonlinear function is given in (\ref{equ:hat_ff}). The average mean-square errors for node $2$ of this uncertain case are provided in Fig.~\ref{f:com_with}. It can be found that in either case, the estimation performance of DREA is better than that of AREA and NVEA. This further shows the superiority of the proposed estimators in terms of handling uncertainties, disturbances and nonlinearities.

\textbf{Part 3:} To better illustrate the effectiveness of the proposed algorithm in dealing with general network structures, the simulation for network (\ref{equ:observed_system}) with different topologies is presented, including a star graph,  a chain graph and a ring graph, as shown in Fig.~\ref{f:comgraph}. In this part, the system model is chosen to be the same as that in \textbf{Part 1}, expect the coupling strengths $\pi_{ij}$. For the star graph, $\pi_{11}=-0.1$, $\pi_{12}=0.1$, $\pi_{21}=0.1$, $\pi_{22}=-0.3$, $\pi_{23}=0.1$, $\pi_{24}=0.1$, $\pi_{32}=0.1$, $\pi_{33}=-0.1$, $\pi_{42}=0.1$ and $\pi_{44}=-0.1$. For the ring graph, $\pi_{11}=-0.2$, $\pi_{12}=0.1$, $\pi_{14}=0.1$, $\pi_{21}=0.1$, $\pi_{22}=-0.1$, $\pi_{32}=0.1$, $\pi_{33}=-0.2$, $\pi_{34}=0.1$, $\pi_{43}=0.1$ and $\pi_{44}=-0.1$. For the chain graph, $\pi_{21}=0.1$, $\pi_{22}=-0.1$, $\pi_{32}=0.1$, $\pi_{33}=-0.1$, $\pi_{43}=0.1$ and $\pi_{44}=-0.1$. Then, without losing the generality, take node 2 for example, of which the average mean-square errors (dB: $-10 \times \lg(MSE_{k,i}$)) obtained by the proposed algorithm is shown in Fig.~\ref{f:graphcom}. It can be seen that the errors of node $2$ with different topologies are similar, which indicates that the proposed algorithm  can ensure the estimation performance under different network topologies.

From the above three aspects, the effectiveness of the theoretical results in Section \ref{sec3} is well confirmed.

\section{Conclusions} \label{sec5}
In this paper, a state estimator is proposed to estimate the states of a class of uncertain nonlinear networks. The uncertainties, disturbances and nonlinearities in system dynamics are modeled under a unified framework. By using the regularized least-squares estimating and a decoupling technique, the estimator gains are designed by solving a distributed optimization problem. Differing from the LMIs-based approach and the variance-constrained approach, the feasibility of the estimators and the boundedness of MSE are guaranteed with an easily checked condition. The good estimation performance of the proposed estimation algorithm are illustrated by comparative simulations.

\appendices

\section{Derivation of Algorithm \ref{algorithm2}} \label{proof1}
First, the state estimation problem for the uncertain nonlinear networks (\ref{equ:observed_system}) is investigated in the centralized form as a preliminary study. By designing a priori estimate $\overline{x}_{k,i}$ evolving as (\ref{equ:overline_x}), the priori estimation error is defined as $\overline{e}_{k,i} = x_{k,i} - \overline{x}_{k,i}$. From (\ref{equ:observed_system}), (\ref{equ:overline_x}), (\ref{equ:nonlinear}) and (\ref{equ:nonlinearuncertain}), the dynamics of $\overline{e}_{k+1,i}$ are obtained as
\begin{align} \label{equ:overline_e}
 \overline{e}_{k+1,i}
 = & ( F_{k,i} + E_{2,k,i} \Delta_{2,k,i} + E_{1,k,i} \Delta_{1,k,i}) e_{k,i}  + \omega_{k,i} \notag \\
   & + E_{1,k,i} \Delta_{1,k,i} \hat{x}_{k,i}
 + a \sum_{j \in \mathcal{N}_i} \pi_{ij} G  e_{k,j}.
\end{align}

Then, borrowing the reformulation idea for the standard Kalman filtering problem discussed in \cite{sayed2001framework}, the optimization problem (\ref{equ:optimal_uncertain1}) for node $i$ in complex networks (\ref{equ:observed_system}) is considered. Substituting (\ref{equ:overline_e})  into (\ref{equ:J11}) yields
\begin{align}   \label{equ:J12}
    J_{1,i}
 = & \left ( { \sum_{j \in \mathcal{N}_i}  \| e_{k,j} \|^2_{P_{k,j}^{-1}} } \right ) + \| w_{k,i} \|^2_{Q_{k,i}^{-1}}  +  \| -b_{k+1,i} + H_{i} \omega_{k,i} \notag \\
  &  + H_{i} E_{k,i} \Delta_{k,i} [I_n \ I_n]^T e_{k,i}  + H_{i} E_{1,k,i} \Delta_{1,k,i} \hat{x}_{k,i} \notag \\
   &  + H_{i} F_{k,i} e_{k,i}
 + a H_{i} \sum_{j \in \mathcal{N}_i} \pi_{ij} G  e_{k,j}  \|^2_{R_{k+1,i}^{-1}} \notag \\
 = &  J_{1,i}(e_{k,j}, \ w_{k,i}, \ j \in \mathcal{N}_i),
\end{align}
where
\begin{align}
e_{k,i} & = x_{k,i} - \hat{x}_{k,i},  & b_{k+1,i}  =  y_{k+1,i} - H_{i}  \overline{x}_{k+1,i},  \notag \\
E_{k,i} & = [E_{1,k,i}, \ E_{2,k,i}],  & \Delta_{k,i} = \text{diag}[\Delta_{1,k,i}, \ \Delta_{2,k,i}].
\notag
\end{align}

Furthermore, $J_{1,i}$ can be rewritten in the compact form as follows:
\begin{align}   \label{equ:J13}
  J_{1,i}(\eta_{k,i})
 =  & \| \eta_{k,i} \|^2_{S_{k,i}}  + \| A_{k,i} \eta_{k,i} - b_{k+1,i}  \notag \\
  &  +  M_{k,i} \Delta_{k,i} (E_{a,k,i} \eta_{k,i} - E_{b,k,i}) \|^2_{T_{k,i}},
\end{align}
where
\begin{align}
\eta_{k,i} & = [e_{k,1}^T, \ \ldots, \ e_{k,N}^T, \ w_{k,i}^T ]^T ,   \notag \\
 A_{k,i} & = H_{i}[ a w_{i1} G, \ \ldots, \  a w_{i(i-1)} G,  \notag \\
 & \qquad a w_{ii} G + F_{k,i},  a w_{i(i+1)} G,  \ \ldots, \   a w_{iN} G, \ I_n ] ,   \notag \\
 S_{k,i} & =  \text{diag} [P_{k,1}^{-1},  \ \ldots, \  P_{k,N}^{-1}, \ Q_{k,i}^{-1} ] , \notag \\
 E_{a,k,i} & =   [I_n \ I_n]^T  [0_{n \times n(i-1)}, \ I_{n}, \ 0_{n \times n(N+1-i)} ] , \notag \\
 T_{k,i} & = R_{k+1,i}^{-1},  \qquad \qquad \qquad M_{k,i}  = H_{i}  E_{k,i} , \notag \\
 E_{b,k,i} & = - [I_n \ 0]^T \hat{x}_{k,i} . \notag
\end{align}

Taking uncertainties in $J_{1,i}(\eta_{k,i})$ into consideration, the optimization problem (\ref{equ:optimal_uncertain1}) can be reformulated as
\begin{align}   \label{equ:optimal_uncertain3}
  \min_{\eta_{k,i}} \max_{\| \Delta_{k,i} \|_2 \leq 1} J_{1,i}(\eta_{k,i}).
\end{align}
The optimal value of $\eta_{k,i}$ in (\ref{equ:optimal_uncertain3}) is derived as
\begin{align} \label{equ:eta}
 \eta_{k,i}^{opt} = & [ \hat{S}_{k,i} + A_{k,i}^T \hat{T}_{k,i} A_{k,i}]^{-1} [ A_{k,i}^T \hat{T}_{k,i} b_{k+1,i}   + \hat{\lambda}_{k,i} E_{a,k,i}^T E_{b,k,i} ],
\end{align}
where $ \hat{S}_{k,i} =  S_{k,i} + \hat{\lambda}_{k,i} E_{a,k,i}^T E_{a,k,i} $, $ \hat{T}_{k,i} =  ( T_{k,i}^{-1} - \hat{\lambda}_{k,i}^{-1} M_{k,i} M_{k,i}^{T}  )^{-1} $,
and the scalar $\hat{\lambda}_{k,i}$ is derived similarly to that discussed in \cite{sayed2001framework}.

%By adopting the designing ideas of the standard Kalman filter based on the regularized least-squares approach, the state estimate for node $i$ in networks (\ref{equ:observed_system}) at step $k+1$  is given as
%\begin{align} \label{equ:estimate1}
% \hat{x}_{k+1,i}^{c} = &  \overline{x}_{k+1,i} +  [ a w_{i1} G, \ \ldots, \ a w_{i(i-1)} G, \notag \\
% &  F_{k,i} + a w_{ii} G, \ a w_{i(i+1)} G, \ \ldots, \ a w_{iN} G, \ I] \eta_{k,i}^{opt},
%\end{align}
%In particular, according to the structures of $\hat{S}_{k,i}$ and $\hat{T}_{k,i}$, the approximate covariance matrix for node $i$ at step $k+1$  can be directly derived as
%\begin{align} \label{equ:P1}
%P_{k+1,i}  = ( \breve{P}_{k+1,i}^{-1} + H_{i}^T \hat{R}_{k+1,i}^{-1} H_{i} + 2 \hat{\lambda}_{k,i} I_n )^{-1},
%\end{align}
%with
%\begin{align}
%\breve{P}_{k+1,i}  = &  (F_{k,i} + a w_{ii} G)  P_{k,i} (F_{k,i} + a w_{ii} G)^T  + Q_{k,i}   \notag \\
%& + \sum_{j \in \mathcal{N}_i, j \neq i}  a^2 \pi_{ij}^2 G  P_{k,j} G^T ,  \notag \\
%\hat{R}_{k+1,i}   = & R_{k+1,i} - \hat{\lambda}_{k,i}^{-1} H_{i}  E_{k,i} E_{k,i}^{T}  H_{i}^{T} . \notag
%\end{align}

In the following, the distributed algorithm is developed. According to (\ref{equ:eta}), in order to design the distributed estimator, one needs to diagonalize $A_{k,i}^T \hat{T}_{k,i} A_{k,i}$, so as to relieve the coupling relations between $e_{k,i}$ and $e_{k,j}$ in $J_{1,i}(e_{k,j}, \ w_{k,i}, \ j \in \mathcal{N}_i)$.

On the basis of (\ref{equ:J12}), by employing Lemma \ref{lemma2}, one has
\begin{align}   % \label{equ:J21}
 &J_{1,i}(e_{k,j}, \ w_{k,i}, \ j \in \mathcal{N}_i)  \notag \\
 \leq &  \sum_{j \in \mathcal{N}_i}  \| e_{k,j} \|^2_{P_{k,j}^{-1}}  + \| w_{k,i} \|^2_{Q_{k,i}^{-1}}  + (1+a) \| - (1 - \alpha_{k,i}) b_{k+1,i} \notag \\
  & + H_{i} ( F_{k,i} + a \pi_{ii} G_i ) e_{k,i} + H_{i} E_{k,i} \Delta_{k,i} [I \ I]^T e_{k,i} \notag \\
   &  + H_{i} E_{1,k,i} \Delta_{1,k,i} \hat{x}_{k,i} + H_{i} w_{k,i} \|^2_{R_{k+1,i}^{-1} }
  \notag \\
  &   + (a+a^2) \|- \alpha_{k,i} b_{k+1,i}+  H_{i} \sum_{j \in \mathcal{N}_i, j \neq i} \pi_{ij} G  e_{k,j}  \|^2_{R_{k+1,i}^{-1}} , \notag
\end{align}
where $\alpha_{k,i} \in [0$, $1]$ is a scalar to be designed.

Then, according to Lemma \ref{lemma3}, the last term in the above inequality satisfies
\begin{align}
& (a+a^2) \| - \alpha_{k,i} b_{k+1,i} + H_{i} \sum_{j \in \mathcal{N}_i, j \neq i} \pi_{ij} G  e_{k,j}  \|^2_{R_{k+1,i}^{-1} } \notag \\
 = & (a+a^2) \| \sum_{j \in \mathcal{N}_i, j \neq i} \pi_{ij} ( H_{i}  G  e_{k,j} - \hat{\alpha}_{k,i} b_{k+1,i} ) \|^2_{R_{k+1,i}^{-1} } \notag \\
 \leq & \sum_{j \in \mathcal{N}_i, j \neq i}  \|  H_{i}  G  e_{k,j} - \hat{\alpha}_{k,i} b_{k+1,i} \|^2_{Z_{k,ij}} , \notag
\end{align}
where $Z_{k,ij}= \pi_{ij}^2(a+a^2)  (N-1) R_{k+1,i}^{-1}$ and $\hat{\alpha}_{k,i}={\alpha_{k,i}}/(\sum_{j \in \mathcal{N}_i, j \neq i} \pi_{ij})$.

Next, from the above two inequalities, one obtains
\begin{align}  \label{equ:J21}
 & J_{1,i}(e_{k,j}, \ w_{k,i}, \ j \in \mathcal{N}_i)  \notag \\
 \leq &  \sum_{j \in \mathcal{N}_i}  \| e_{k,j} \|^2_{P_{k,j}^{-1}}  + \| w_{k,i} \|^2_{Q_{k,i}^{-1}}  + (1+a) \| - (1 - \alpha_{k,i}) b_{k+1,i} \notag \\
  & + H_{i} ( F_{k,i} + a \pi_{ii} G_i ) e_{k,i} + H_{i} E_{k,i} \Delta_{k,i} [I \ I]^T e_{k,i} \notag \\
   &  + H_{i} E_{1,k,i} \Delta_{1,k,i} \hat{x}_{k,i} + H_{i} w_{k,i} \|^2_{R_{k+1,i}^{-1} }
  \notag \\
  &  + \sum_{j \in \mathcal{N}_i, j \neq i}  \|  H_{i}  G  e_{k,j} - \hat{\alpha}_{k,i} b_{k+1,i} \|^2_{Z_{k,ij}}   \notag \\
  \triangleq & J_{2,i}(e_{k,j}, \ w_{k,i}, \ j \in \mathcal{N}_i).
\end{align}

Like (\ref{equ:J13}), one can rewrite $J_{2,i}$ in a compact form as
\begin{align}  \label{equ:J22}
 & J_{2,i}(e_{k,j}, \ w_{k,i}, \ j \in \mathcal{N}_i)  \notag \\
 = & J_{2,i}(\overline{\eta}_{k,i}, \ e_{k,j}( j \in \mathcal{N}_i, \ j \neq i))  \notag \\
 = &   \| \overline{\eta}_{k,i} \|^2_{\overline{S}_{k,i}}  +  \| \overline{A}_{k,i} \overline{\eta}_{k,i} - (1 - \alpha_{k,i}) b_{k+1,i}  +  \overline{M}_{k,i}  \Delta_{k,i} \notag \\
  & \times ( \overline{E}_{a,k,i} \overline{\eta}_{k,i} - \overline{E}_{b,k,i})  \|^2_{\overline{T}_{k,i}}   + \sum_{j \in \mathcal{N}_i, j \neq i} \Big{(}  \| e_{k,j} \|^2_{P_{k,j}^{-1}}   \notag \\
  & +  \|  H_{i}  G  e_{k,j} - \hat{\alpha}_{k,i}  b_{k+1,i} \|^2_{Z_{k,ij}}  \Big{)},
\end{align}
where
\begin{align}
\overline{\eta}_{k,i} & = [e_{k,i},  \ w_{k,i} ],  &    \overline{A}_{k,i} = H_{i}[ a \pi_{ii} G + F_{k,i}, \ I_n ],    \notag \\
 \overline{S}_{k,i} & =  \text{diag} [P_{k,i}^{-1}, \ Q_{k,i}^{-1} ],
 & \overline{M}_{k,i}  = H_{i}  E_{k,i},  \notag \\
 \overline{T}_{k,i} & = (1+a) R_{k+1,i}^{-1},  & \overline{E}_{a,k,i} =   [I_n \ I_n]^T [I_n \ 0_n], \notag \\
 \overline{E}_{b,k,i} & = - [I_n \ 0_n]^T \hat{x}_{k,i} . \notag
\end{align}

The above steps convert a centralized algorithm to a distributed one. In particular, the optimization problem (\ref{equ:optimal_uncertain3}) about $J_{1,i}$ is translated to the following one about $J_{2,i}$:
\begin{align}   \label{equ:optimal_uncertain4}
  \min_{\stackrel{\overline{\eta}_{k,i}} {e_{k,j} }} \max_{\| \Delta_{k,i} \|_2 \leq 1} J_{2,i}(\overline{\eta}_{k,i}, \ e_{k,j}( j \in \mathcal{N}_i, \ j \neq i)).
\end{align}
Note that, by the above conversion process from $J_{1,i}$ to $J_{2,i}$, the components of the cost function are divided into two parts: 1) node $i$'s own estimate $\overline{\eta}_{k,i}$;  2) node $i$'s neighbors' estimates $e_{k,j}, \ j \in \mathcal{N}_i, \ j \neq i$. Since these two parts have been decoupled, $\overline{\eta}_{k,i}$ and $e_{k,j}$ can be optimized respectively, i.e.,
\begin{align}
 \overline{\eta}_{k,i}^{opt} = \arg \min_{\overline{\eta}_{k,i}} \max_{\| \Delta_{k,i} \|_2 \leq 1} J_{21,i}(\overline{\eta}_{k,i}), \notag
\end{align}
and
\begin{align}
  e_{k,ij} ^{opt} = & \arg \min_{ e_{k,j} } \max_{\| \Delta_{k,i} \|_2 \leq 1}  J_{22,i}(e_{k,j}), \notag \\
   & \ j \in \mathcal{N}_i,  \ j \neq i \notag
\end{align}
where
\begin{align} \label{equ:J21_ETA}
 J_{21,i}(\overline{\eta}_{k,i}) = & \| \overline{\eta}_{k,i} \|^2_{\overline{S}_{k,i}}  +  \| \overline{A}_{k,i} \overline{\eta}_{k,i} - (1 - \alpha_{k,i}) b_{k+1,i}  \notag \\
  &  +  \overline{M}_{k,i}  \Delta_{k,i}  ( \overline{E}_{a,k,i} \overline{\eta}_{k,i} - \overline{E}_{b,k,i})  \|^2_{\overline{T}_{k,i}},
\end{align}
and
\begin{align} \label{equ:J21_eij}
  J_{22,i}(e_{k,j})  = \| e_{k,j} \|^2_{P_{k,j}^{-1}} +  \|  H_{i}  G  e_{k,j} - \hat{\alpha}_{k,i}  b_{k+1,i} \|^2_{Z_{k,ij}}.
\end{align}

By using the regularized least-squares strategies with and without system matrix uncertainties respectively, one can formulate $\overline{\eta}_{k,i}^{opt}$ and $e_{k,ij}^{opt}$ as
\begin{align}  \label{equ:overline_eta}
 \overline{\eta}_{k,i}^{opt} = & [ \hat{\overline{S}}_{k,i} + \overline{A}_{k,i}^T \hat{\overline{T}}_{k,i} \overline{A}_{k,i}]^{-1} [ (1 - \alpha_{k,i} ) \overline{A}_{k,i}^T \hat{\overline{T}}_{k,i} b_{k+1,i} \notag \\
  & + \overline{\lambda}_{k,i}  \overline{E}_{a,k,i}^T \overline{E}_{b,k,i} ],
\end{align}
and
\begin{align} \label{equ:e_ij}
  e_{k,ij} ^{opt} = & \hat{\alpha}_{k,i} [ P_{k,j}^{-1} +  G^T H_{i}^T Z_{k,ij} H_{i}  G]^{-1}   G^T H_{i}^T Z_{k,ij} b_{k+1,i},
\end{align}
where the parameter $\overline{\lambda}_{k,i}$ is a scalar to be optimized later, and
\begin{align}   \label{equ:hat_S}
 \hat{\overline{S}}_{k,i} = &  \overline{S}_{k,i} + \overline{\lambda}_{k,i} \overline{E}_{a,k,i}^T \overline{E}_{a,k,i} ,  \\
 \label{equ:hat_T} \hat{\overline{T}}_{k,i} = &  ( \overline{T}_{k,i}^{-1} - \overline{\lambda}_{k,i}^{-1} \overline{M}_{k,i} \overline{M}_{k,i}^{T}  )^{-1} = \hat{R}_{k+1,i}^{-1}.
\end{align}

Similarly to the centralized algorithm, based on the structures of $\hat{\overline{S}}_{k,i}$ and $\hat{\overline{T}}_{k,i}$, the approximate covariance matrix $P_{k+1,i}$ is derived as (\ref{equ:P2}).
Now, one can propose the state estimator as
\begin{align} \label{equ:estimate2}
  \hat{x}_{k+1,i} = &  \overline{x}_{k+1,i} + [F_{k,i} + a \pi_{ii} G, \ I] \overline{\eta}_{k,i}^{opt}   +  \sum_{j \in \mathcal{N}_i, j \neq i} a \pi_{ij} G e_{k,ij} ^{opt},
\end{align}
where $\overline{x}_{k+1,i}$ is given in (\ref{equ:overline_xd}).
\vspace{6pt}

Moreover, since $\overline{\eta}_{k,i}^{opt}$ is augmented, to further reduce the computational cost, $e_{k,i}^{opt}$ and $w_{k,i}^{opt}$, the components of $\overline{\eta}_{k,i}^{opt}$, need to be solved in the decoupling form. Alternatively, $[F_{k,i} + a \pi_{ii} G, \ I] \overline{\eta}_{k,i}^{opt}$, the second term in (\ref{equ:estimate2}), needs to be computed in a non-augmented manner.

After some tedious computations,  it follows from (\ref{equ:overline_eta}) and Lemma \ref{lemma1} that
\begin{small}
\begin{align}
 & [F_{k,i} + a \pi_{ii} G, \ I] \overline{\eta}_{k,i}^{opt} \notag \\
 = & (1 - \alpha_{k,i} ) [F_{k,i} + a \pi_{ii} G, \ I] ( \hat{\overline{S}}_{k,i} + \overline{A}_{k,i}^T \hat{\overline{T}}_{k,i} \overline{A}_{k,i})^{-1}    \overline{A}_{k,i}^T \hat{\overline{T}}_{k,i} b_{k+1,i}   \notag \\
  & +  \overline{\lambda}_{k,i}  [F_{k,i} + a \pi_{ii} G, \ I] [ \hat{\overline{S}}_{k,i} + \overline{A}_{k,i}^T \hat{\overline{T}}_{k,i} \overline{A}_{k,i}]^{-1}   \overline{E}_{a,k,i}^T \overline{E}_{b,k,i} ] \notag \\
 = & (1 - \alpha_{k,i} ) ( \breve{P}_{k+1,i}^{-1} + H_{i}^{T} \hat{\overline{T}}_{k,i} H_{i}  +  2\overline{\lambda}_{k,i} I_n )^{-1} H_{i}^{T}  \hat{\overline{T}}_{k,i} b_{k+1,i} \notag \\
  &  +  \overline{\lambda}_{k,i} [ ( \breve{P}_{k+1,i}^{-1} + H_{i}^{T} \hat{\overline{T}}_{k,i} H_{i}  +  2 \overline{\lambda}_{k,i} I_n )^{-1}  H_{i}^{T} \hat{\overline{T}}_{k,i} H_{i} - I] \notag \\
  & \times (F_{k,i} + a \pi_{ii} G)  P_{k,i} \hat{x}_{k,i} ,\notag \\
  = & (1 - \alpha_{k,i} ) K_{1,k+1,i} b_{k+1,i}  +  \overline{\lambda}_{k,i} [ K_{1,k+1,i} H_{i} - I] (F_{k,i} + a \pi_{ii} G) \notag \\
   & \times P_{k,i} \hat{x}_{k,i} ,\notag
  \end{align}
  \end{small}
where $\breve{P}_{k+1,i}^{-1}=(F_{k,i} + a \pi_{ii} G) P_{k,i} (F_{k,i} + a \pi_{ii} G)^T + Q_{k,i} $.
Subsequently, the estimator (\ref{equ:estimate2}) can be re-written as (\ref{equ:estimate3}).

\section{Proof of Theorem \ref{thm3}}\label{proof2}
Generally, the parameter $\overline{\lambda}_{k,i}$ can be obtained by
\begin{align}
 & \overline{\lambda}_{k,i}^{opt} (\alpha_{k,i})=  \arg \min_{ \lambda_{k,i} \ge \| H^{T} T H \|} G_1(\lambda_{k,i}, \alpha_{k,i}) ,\notag
\end{align}
where
\begin{align}
  & G_1(\overline{\lambda}_{k,i}, \alpha_{k,i}) \notag \\
  = & \overline{\lambda}_{k,i} \overline{E}_{b,k,i}^T \overline{E}_{b,k,i}
 - 2 \overline{\lambda}_{k,i} \overline{E}_{b,k,i}^T \overline{E}_{a,k,i} \overline{\eta}_{k,i} + \overline{\eta}_{k,i}^T \hat{\overline{S}}_{k,i} \overline{\eta}_{k,i}  \notag \\
 & + \| \overline{A}_{k,i} \overline{\eta}_{k,i} - (1 - \alpha_{k,i}) b_{k+1,i} \|^2_{\hat{\overline{T}}_{k,i}}.  \notag
\end{align}

Then, after some calculations, it can be verified that $J_{21,i}(\overline{\eta}_{k,i})$ in (\ref{equ:J21_ETA}) satisfies
\begin{align}
 & J_{21,i}(\overline{\eta}_{k,i})  \notag \\
 \leq & G_1( \overline{\lambda}^{opt}_{k,i}(\alpha_{k,i}), \alpha_{k,i}) \notag \\
 =  & \overline{\lambda}^{opt}_{k,i}(\alpha_{k,i}) \overline{E}_{b,k,i}^T \overline{E}_{b,k,i}
 - 2 \overline{\lambda}^{opt}_{k,i} (\alpha_{k,i}) \overline{E}_{b,k,i}^T \overline{E}_{a,k,i} \overline{\eta}_{k,i}  \notag \\
 & + \overline{\eta}_{k,i}^T \hat{\overline{S}}_{k,i} \overline{\eta}_{k,i} + \| \overline{A}_{k,i} \overline{\eta}_{k,i} - (1 - \alpha_{k,i}) b_{k+1,i} \|^2_{\hat{\overline{T}}_{k,i}}  \notag \\
 \triangleq &  \hat{J}_{21,i}(\overline{\eta}_{k,i},\overline{\lambda}^{opt}_{k,i}(\alpha_{k,i}), \alpha_{k,i}). \notag
\end{align}

Equivalently, it follows from (\ref{equ:J22})  that
\begin{align} \label{equ:hat_J2}
  & J_{2,i}(\overline{\eta}_{k,i}, \ e_{k,j}( j \in \mathcal{N}_i, j \neq i))   \notag \\
 = &  J_{21,i}(\overline{\eta}_{k,i}) + \sum_{j \in \mathcal{N}_i, j \neq i}  J_{22,i}(e_{k,j}) \notag \\
  \leq &  \hat{J}_{21,i}(\overline{\eta}_{k,i},\overline{\lambda}^{opt}_{k,i}(\alpha_{k,i}), \alpha_{k,i})  + \sum_{j \in \mathcal{N}_i, j \neq i}  J_{22,i}(e_{k,j}) \notag \\
  \triangleq &  \hat{J}_{2,i}(\overline{\eta}_{k,i}, e_{k,j}, \overline{\lambda}^{opt}_{k,i}(\alpha_{k,i}), \alpha_{k,i}).
\end{align}

Note that the scalars $\alpha_{k,i}$ and $\overline{\lambda}_{k,i}$ can be optimized by the same cost function, as
\begin{align}
 & \alpha_{k,i}^{opt} =  \arg \min_{ \alpha_{k,i} \in [0, \ 1] } \hat{J}_{2,i}(\overline{\eta}_{k,i},e_{k,j},\overline{\lambda}_{k,i}, \alpha_{k,i}) ,\notag
\end{align}
and
\begin{align}
\overline{\lambda}_{k,i}^{opt} =  &  \arg \min_{ \overline{\lambda}_{k,i} \ge \| H^{T} T H \|_2} G_1(\overline{\lambda}_{k,i}, \alpha_{k,i}) \notag \\
=  & \arg \min_{ \overline{\lambda}_{k,i} \ge \| H^{T} T H \|_2}  \hat{J}_{2,i}(\overline{\eta}_{k,i},e_{k,j},\overline{\lambda}_{k,i}, \alpha_{k,i}).  \notag
\end{align}

Next, substituting $\overline{\eta}_{k,i}^{opt}$ and $e_{k,ij}^{opt}$ in (\ref{equ:overline_eta}) and (\ref{equ:e_ij}) into $\hat{J}_{2,i}(\overline{\eta}_{k,i},e_{k,j},\overline{\lambda}_{k,i}, \alpha_{k,i})$ yields
\begin{align} \label{equ:J_Alpha_L}
 & \hat{J}_{2,i}(\overline{\lambda}_{k,i}, \alpha_{k,i}) \notag \\
   =  & \overline{\lambda}_{k,i} \overline{E}_{b,k,i}^T \overline{E}_{b,k,i} + {\overline{\eta}_{k,i}^{opt}}^T \hat{\overline{S}}_{k,i} \overline{\eta}_{k,i}^{opt}
 - 2 \overline{\lambda}_{k,i} \overline{E}_{b,k,i}^T \overline{E}_{a,k,i} \overline{\eta}_{k,i}^{opt}  \notag \\
  &  + \sum_{j \in \mathcal{N}_i, j \neq i} ( \| e_{k,ij}^{opt} \|^2_{P_{k,j}^{-1}}   +  \|  H_{i}  G  e_{k,ij}^{opt} - \hat{\alpha}_{k,i}  b_{k+1,i} \|^2_{Z_{k,ij}} )  \notag \\
  & + \| \overline{A}_{k,i} \overline{\eta}_{k,i}^{opt} - (1 - \alpha_{k,i}) b_{k+1,i} \|^2_{\hat{\overline{T}}_{k,i}},
\end{align}
where $\hat{\overline{S}}_{k,i}$ and $\hat{\overline{T}}_{k,i}$ are defined in (\ref{equ:hat_S}) and (\ref{equ:hat_T}), respectively, and $\hat{\alpha}_{k,i}={\alpha_{k,i}}/(\sum_{j \in \mathcal{N}_i, j \neq i} \pi_{ij})$. Other matrices are given in (\ref{equ:J22}).
Thus, the parameters $\overline{\lambda}_{k,i}$ and $\alpha_{k,i}$ can be optimized by the optimization (\ref{equ:alpha_lambda}).

Thus, the proof of Theorem \ref{thm3} is complete.

\section{Proof of  Corollary \ref{cor1}}\label{proof3}
First, substituting (\ref{equ:overline_eta}) and (\ref{equ:e_ij}) into (\ref{equ:hat_J2}) yields
% \begin{align}
% & \hat{J}_{2,i}(\alpha_{k,i})  \notag \\
% = & \hat{\overline{\lambda}}_{k,i} \overline{E}_{b,k,i} \overline{E}_{b,k,i}^T - 2 \hat{\overline{\lambda}}_{k,i} \overline{E}_{b,k,i}^T \overline{E}_{a,k,i} \overline{\eta}_{k,i}^{opt}  \notag \\
%  & - (1 - \alpha_{k,i}) b_{k+1,i}^T \hat{\overline{T}}_{k,i}  \overline{A}_{k,i} \overline{\eta}_{k,i}^{opt} + (1 - \alpha_{k,i})^2 b_{k+1,i}^T \hat{\overline{T}}_{k,i}  b_{k+1,i}
%  \notag \\
%   & + \sum_{j \in \mathcal{N}_i, j \neq i} - \hat{\alpha}_i b_{k+1,i}^T Z_{k,ij} H_{i} G  e_{k,ij}^{opt} + \hat{\alpha}_i^2 b_{k+1,i}^T Z_{k,ij} b_{k+1,i} \notag
%\end{align}
\begin{align} % \label{equ:hat_J22}
\hat{J}_{2,i}(\alpha_{k,i})
 =   & \alpha_{k,i}^2  \left ({ \Phi_{k,i}^{2} + \sum_{j \in \mathcal{N}_i, j \neq i} \Phi_{k,ij}^{3} }  \right ) - \alpha_{k,i} ( \Phi_{k,i}^{1} + 2  \Phi_{k,i}^{2} ) \notag \\
 & + \Phi_{k,i}^{0} + \Phi_{k,i}^{1} + \Phi_{k,i}^{2}, \notag
 \end{align}
where  $\Phi_{k,i}^{0}$, $\Phi_{k,i}^{1}$, $\Phi_{k,i}^{2}$, and $\Phi_{k,ij}^{3}$ are defined as
\begin{small}
\begin{align} \label{equ:Phi}
\Phi_{k,i}^{0} = & - 2 \overline{\lambda}_{k,i}^2 \overline{E}_{b,k,i}^T \overline{E}_{a,k,i}[ \hat{\overline{S}}_{k,i} + \overline{A}_{k,i}^T \hat{\overline{T}}_{k,i} \overline{A}_{k,i}]^{-1} \overline{E}_{a,k,i}^T \overline{E}_{b,k,i} \notag \\
 & + \overline{\lambda}_{k,i} \overline{E}_{b,k,i}^T \overline{E}_{b,k,i} , \notag \\
\Phi_{k,i}^{1} = & - 3 \overline{\lambda}_{k,i} \overline{E}_{b,k,i}^T \overline{E}_{a,k,i}[ \hat{\overline{S}}_{k,i} + \overline{A}_{k,i}^T \hat{\overline{T}}_{k,i} \overline{A}_{k,i}]^{-1} \overline{A}_{k,i}^T \hat{\overline{T}}_{k,i} b_{k+1,i}  ,\notag \\
\Phi_{k,i}^{2} = &   b_{k+1,i}^T (\hat{\overline{T}}_{k,i}^{-1} + \overline{A}_{k,i} \hat{\overline{S}}_{k,i}^{-1} \overline{A}_{k,i}^T )^{-1} b_{k+1,i} ,\notag \\
\Phi_{k,ij}^{3} = & b_{k+1,i}^T ( Z_{k,ij}^{-1} +  H_{i}  G P_{k,j} G^T H_{i}^T )^{-1} b_{k+1,i} \Big / \bigg ( \sum_{j \in \mathcal{N}_i, j \neq i} \pi_{ij} \bigg ).
\end{align}
\end{small}

Then, since $\Phi_{k,i}^{0}$, $\Phi_{k,i}^{1}$, $\Phi_{k,i}^{2}$, and $\Phi_{k,ij}^{3}$ are scalars, after some simple calculations, one has
\begin{align}
\alpha_{k,i}^{opt} & = \arg \min_{\alpha_{k,i}} \hat{J}_{2,i}(\alpha_{k,i}) \notag \\
& = (\Phi_{k,i}^{1} + 2  \Phi_{k,i}^{2} ) / \left ({ 2 \Phi_{k,i}^{2} + 2 \sum_{j \in \mathcal{N}_i, j \neq i} \Phi_{k,ij}^{3} }  \right ) .\notag
\end{align}
Since $\alpha_{k,i}^{opt} \in [0$, $1]$,  $\alpha_{k,i}^{opt}$ can be  solved by (\ref{equ:alpha_opt}).

Thus, the proof of Corollary \ref{cor1} is complete.

\section{Proof of Theorem \ref{thm4}}\label{proof4}

The proof of Theorem \ref{thm4} is divided into two parts: 1) $ P_{k,i}$ is uniformly upper bounded;  2) $ P_{k,i}$ is uniformly lower bounded.

\textbf{Part 1:} By the definition of $P_{k+1,i}$ in (\ref{equ:P2}) and Lemma \ref{lemma1}, one has
\begin{align}
P_{k+1,i}  \leq & (1+a)( \breve{P}_{k+1,i}^{-1}  + H_{i}^{T} \hat{R}_{k+1,i}^{-1} H_{i} + 2 \overline{\lambda}_{k,i} I_n )^{-1} \notag \\
 & +  (a+a^2)  (N-1) \sum_{j \in \mathcal{N}_i, j \neq i}  \pi_{ij}^2  G P_{k,j} G^T. \notag
\end{align}

Now, define an augmented recursive matrix $P_{k+1}= \text{diag}  [P_{k+1,1}$, $\ldots$, $P_{k+1,N}]$. It follows from the above inequality that
\begin{align} \label{equ:PP}
P_{k+1}  \leq (1+a) ( \breve{P}_{k+1}^{-1}  + H^{T} \hat{R}_{k+1}^{-1} H + 2 \overline{\lambda}_{k} )^{-1}   + \hat{G}_{k} P_{k}^N \hat{G}_{k}^T,
\end{align}
with
\begin{align}
\label{equ:breve_PP}  \breve{P}_{k+1} = & F_{k}  P_{k} F_{k}^T + Q_{k},
\end{align}
where $H=\text{diag}  [H_{1}$, $\ldots$, $H_{N}]$, $\hat{R}_{k+1}=\text{diag}  [\hat{R}_{k+1,1}$, $\ldots$, $\hat{R}_{k+1,N}]$, $\hat{G}_{k}=\text{diag}  [G^{1}$, $\ldots$, $G^{N}]$ with $G^{i} = [\pi_{i1} G$, $\ldots$, $\pi_{i(i-1)} G$, $0$, $\pi_{i(i+1)} G$, $\ldots$, $\pi_{iN} G]$, $P_{k}^N=(a+a^2)  (N-1) \times \text{diag}  [P_{k}$, $\ldots$, $P_{k}]_{N}$, $F_{k}=\text{diag}  [F_{k,1} + a \pi_{11} G$, $\ldots$, $F_{k,N} + a \pi_{NN} G]$,
$Q_{k+1}=\text{diag}  [Q_{k+1,1}$, $\ldots$, $Q_{k+1,N}]$, $\overline{\lambda}_{k} =\text{diag}  [\overline{\lambda}_{k,1} I_n$, $\ldots$, $\overline{\lambda}_{k,N} I_n]$.

According to (\ref{equ:breve_PP}), if Assumptions \ref{asm1} and  \ref{asm3} hold, one has
\begin{align}  \label{equ:inverse_breve_PP}
\breve{P}_{k+1}^{-1} = & (F_{k}  \hat{P}_{k} F_{k}^T + Q_{k})^{-1}
\ge \beta_1 {F_{k}^T}^{-1}  \hat{P}_{k}^{-1} F_{k}^{-1},
\end{align}
with the positive scalar $\beta_1 \in (0,1)$.
Next, considering the inverse of $M_{k+1}= (1+a) ( \breve{P}_{k+1}^{-1}  + H^{T} \hat{R}_{k+1}^{-1} H + 2 \overline{\lambda}_{k} )^{-1} $, it follows from (\ref{equ:PP}) and (\ref{equ:inverse_breve_PP}) that
\begin{align}
 M_{k+1}^{-1}
 \ge  (1+a)^{-1} ( \breve{P}_{k+1}^{-1}  + H^{T} \hat{R}_{k+1}^{-1} H + 2 \overline{\lambda}_{k})
 >  2 (1+a)^{-1} \overline{\lambda}_{k}. \notag
\end{align}
Besides, $M_{k+1}^{-1}$ also satisfies
\begin{align}
M_{k+1}^{-1} \ge & (1+a)^{-1} ( \breve{P}_{k+1}^{-1}  + H^{T} \hat{R}_{k+1}^{-1} H + 2 \overline{\lambda}_{k}) \notag  \\
 > & (1+a)^{-1} [\beta_1 (F_{k}^T)^{-1}  P_{k}^{-1} F_{k}^{-1} + H^{T} \hat{R}_{k+1}^{-1} H] \notag  \\
 \ge & (1+a)^{-1} [\beta_1 \beta_2 (F_{k}^T)^{-1}  M_{k}^{-1} F_{k}^{-1} + H^{T} \hat{R}_{k+1}^{-1} H] \notag  \\
 > & (1+a)^{-2} \beta_1^2 \beta_2^2 (F_{k}^T)^{-1} (F_{k-1}^T)^{-1}  M_{k-1}^{-1} F_{k-1}^{-1} F_{k}^{-1}  \notag \\
 & +  (1+a)^{-2} \beta_1 \beta_2  (F_{k}^T)^{-1} H_{k}^{T} \hat{R}_{k}^{-1}  F_{k}^{-1} H_{k}  \notag \\
 & + (1+a)^{-1}  H^{T} \hat{R}_{k+1}^{-1} H \notag  \\
 > & \cdots \notag \\
 > & (1+a)^{-\overline{N}} \beta_1^{\overline{N}} \beta_2^{\overline{N}} (\Phi_{k,k+1-\overline{N}}^T)^{-1}  M_{k+1-\overline{N}}^{-1} \Phi_{k,k+1-\overline{N}}^{-1}  \notag \\
 & + \sum_{h=0}^{\overline{N}-1} \{ (1+a)^{-h-1} \beta_1^{h} \beta_2^{h} (\Phi_{k,k+1-h}^T)^{-1}   H_{k+1-h}^{T} \notag \\
 & \times \hat{R}_{k+1-h}^{-1} H_{k+1-h} \Phi_{k,k+1-h}^{-1} \} \notag \\
 > & \sum_{h=0}^{\overline{N}-1} \{ (1+a)^{-h-1} \beta_1^{h} \beta_2^{h} (\Phi_{k,k+1-h}^T)^{-1}   H_{k+1-h}^{T} \hat{R}_{k+1-h}^{-1} \notag \\
 & \times H_{k+1-h} \Phi_{k,k+1-h}^{-1} \}, \notag
\end{align}
where $\beta_2$ is a positive scalar. When $\overline{\lambda}_{k,i}=(1+\beta) \breve{\lambda}_{k,i}$ with $\breve{\lambda}_{k,i}$ given in (\ref{equ:breve_lambda}), according to Assumptions {\ref{asm2}} and {\ref{asm3}}, there must exist a scalar $\kappa_2$ such that
\begin{align}
\kappa_2 I_{Nn} \leq & \sum_{h=0}^{\overline{N}-1} \{ (1+a)^{-h-1} \beta_1^{h} \beta_2^{h} (\Phi_{k,k+1-h}^T)^{-1}   H_{k+1-h}^{T} \notag \\
 & \times  \hat{R}_{k+1-h}^{-1} H_{k+1-h} \Phi_{k,k+1-h}^{-1} \} . \notag
\end{align}

Therefore, $M_{k+1}^{-1} > \kappa_3 I_{Nn} \triangleq \max \{ 2 (1+a)^{-1} \overline{\lambda}_{k}, \ \kappa_2 I_{Nn} \}$. From (\ref{equ:PP}), one has
\begin{align}
P_{k+1} -  \hat{G} P_{k}^N \hat{G}^T < \kappa_3^{-1} I_{Nn} , \notag
\end{align}

Then, by pre-multiplying $[I_n$, $\ldots$, $I_n]_N$ and post-multiplying $[I_n$, $\ldots$, $I_n]_N^T$ to the above inequality, it follows that
\begin{align}
  & \kappa_3^{-1} I_{n} \notag \\
  > & \sum_{i=1}^{N} P_{k+1,i} - (a+a^2)  (N-1) \sum_{i=1}^{N}  \sum_{j \in \mathcal{N}_i, j \neq i} \pi_{ij}^2 G P_{k,j} G^T \notag \\
  \ge & \sum_{i=1}^{N} P_{k+1,i} -  (a+a^2)  (N-1)^2 \pi_{m}^2  \sum_{i=1}^{N}    G P_{k,i} G^T \notag \\
  = & \bigg ( \sum_{i=1}^{N} P_{k+1,i} \bigg ) - (a+a^2)  (N-1)^2 \pi_{m}^2 G \bigg ( \sum_{i=1}^{N}    P_{k,i}  \bigg ) G^T , \notag
\end{align}
where $\pi_{m}=\max \{ \pi_{ij}, \ i,j \in [1,\ \ldots,\ N] \}$. Since $ \max \{ \overline{S}_{\lambda}^{Re} ( G ) \} < (a+a^2)^{-\frac{1}{2}}  (N-1)^{-1} \pi_{m}^{-1}$,  one can conclude that  $\sum_{i=1}^{N} P_{k,i}$ is uniformly upper bounded by the well-known Lyapunov method, which further indicates that $P_{k,i}$ is uniformly upper bounded because   $P_{k,i}>0$.
\vspace{6pt}

\textbf{Part 2:} From (\ref{equ:P2}) and (\ref{equ:breve_P2}), if Assumption (\ref{asm3}) holds and $\overline{\lambda}_{k,i}=(1+\beta) \breve{\lambda}_{k,i}$, one has
\begin{align}
 P_{k+1,i}^{-1} \leq & (1+a)^{-1}[ \breve{P}_{k+1}^{-1}  + H^{T} \hat{R}_{k+1}^{-1} H + 2 \overline{\lambda}_{k,i} I_n] \notag \\
 = & (1+a)^{-1} [Q_{k,i}^{-1}  + H^{T} \hat{R}_{k+1}^{-1} H + 2 (1+\beta) \breve{\lambda}_{k,i}I_n] \notag \\
 \leq & (1+a)^{-1} [ \kappa_{\omega,1}^{-1}  + (1+\beta^{-1}) \kappa_{H}^{2} \kappa_{\nu,1}^{-1}  \notag \\
 & + 2 (1+\beta) (\kappa_{E,1}^{2} + \kappa_{E,2}^{2})\kappa_{H}^{2} \kappa_{\nu,1}^{-1} ] I_n, \notag
\end{align}
which means that $P_{k+1,i}$ is uniformly lower bounded.

Thus, the proof of Theorem \ref{thm4} is complete.

\section{Proof of Corollary \ref{cor2}}\label{proof5}

Note that the convergence of $P_{k,i}$ is equivalent to that of $\breve{P}_{k,i}$. Here, the latter is discussed.

First, consider the monotonicity of $\breve{P}_{k,i}$ under Assumption \ref{asm4}.
From (\ref{equ:P2}) and (\ref{equ:breve_P2}), the recursion between $\breve{P}_{k+1,i}$ and $\breve{P}_{k,i}$ is derived as
\begin{align}
\breve{P}_{k+1,i} = &  (F_{i} + a \pi_{ii} G) P_{k,i} (F_{i} + a \pi_{ii} G)^T + Q_{i}  \notag  \\
 = & (1+a) (F_{i} + a \pi_{ii} G) ( \breve{P}_{k,i}^{-1} + H_{i}^{T} \hat{R}_{i}^{-1} H_{i} + 2 \overline{\lambda}_{i} I_n)^{-1} \notag \\
  \times (F_{i} & + a \pi_{ii} G)^T + Q_{i}  +  (a+a^2)(N-1) (F_{i} + a \pi_{ii} G) \notag  \\
\times   \sum_{j \in \mathcal{N}_i, j \neq i} & \pi_{ij}^2  G  ( P_{k,j}^{-1} + \pi_{ij}^2  G^T   H_{i}^T Z_{ij} H_{i} G)^{-1}   G^T (F_{i} + a \pi_{ii} G)^T . \notag
\end{align}

Next, the following proof is divided into two parts: 1) $k=0$; 2) $k\ge 1$.

\textbf{Part 1:} When $k=0$, according to the Initialization in Algorithm \ref{algorithm2}, one has
\begin{align}
\breve{P}_{1,i} \ge & Q_{i} \ge \breve{P}_{0,i}, \ \ \forall  i \in [1, \ \ldots, \ N]  \notag
\end{align}

\textbf{Part 2:} When $k \ge 1$, one has
\begin{align} \label{equ:minus_breve}
& \breve{P}_{k+1,i} - \breve{P}_{k,i}  \notag \\
 = & (1+a) (F_{i} + a \pi_{ii} G)[ ( \breve{P}_{k,i}^{-1} + H_{i}^{T} \hat{R}_{i}^{-1} H_{i} + 2 \overline{\lambda}_{i} I_n)^{-1} \notag \\
 & - ( \breve{P}_{k-1,i}^{-1} + H_{i}^{T} \hat{R}_{i}^{-1} H_{i} + 2 \overline{\lambda}_{i} I_n)^{-1}](F_{i} + a \pi_{ii} G)^T \notag \\
 & +  (a+a^2)(N-1) (F_{i} + a \pi_{ii} G) \notag  \\
& \times \sum_{j \in \mathcal{N}_i, j \neq i}  \pi_{ij}^2  G  [( P_{k,j}^{-1} + \pi_{ij}^2  G^T   H_{i}^T Z_{ij} H_{i} G)^{-1}  \notag \\
-  & ( P_{k-1,j}^{-1} + \pi_{ij}^2  G^T   H_{i}^T Z_{ij} H_{i} G)^{-1} ] G^T (F_{i} + a \pi_{ii} G)^T.
\end{align}
When $\breve{P}_{k,i} > \breve{P}_{k-1,i}$, $\forall  i \in [1, \ \ldots, \ N]$, after some simple calculations, from (\ref{equ:minus_breve}) one can show that $\breve{P}_{k+1,i} > \breve{P}_{k,i}$.

Now, it can be concluded that $\breve{P}_{k,i}$ is a strictly monotonically increasing matrix. By combining it with the boundedness established in Theorem \ref{thm4}, $\breve{P}_{k,i}$ is convergent. Equivalently, $P_{k,i}$ is convergent.

Thus, the proof of Corollary \ref{cor2} is complete.

\section{Proof of Theorem \ref{thm5}}\label{proof6}

Note that $\hat{J}_{2,i}(\alpha_{k,i}=0) \ge  \hat{J}_{2,i}(\alpha_{k,i}^{opt})$. Thus, based on the structure of $\hat{J}_{2,i}$ and the relationship between the Kalman filtering problem and the Regularized Least-squares problem as discussed in \cite{SAYED2002A}, it follows that the estimation performance by Algorithm \ref{algorithm2} with $\alpha_{k,i}^{opt}$ is better than that with $\alpha_{k,i}=0$ in terms of the estimation error covariance. Hence, one only needs to study the case of $\alpha_{k,i}=0$ to analyze the boundedness of MSE.

If $\Delta f_{k,i}=0$, then $\overline{E}_{a,k,i}=[0_n \ I_n]^T [I_n \ 0_n]$ and $\overline{E}_{b,k,i}=[0_n \ 0_n]$ in (\ref{equ:J22}). Besides, when $\alpha_{k,i}=0$ and Assumption $\ref{asm4}$ holds, the estimator (\ref{equ:estimate3}) becomes
\begin{align}
  \hat{x}_{k+1,i} = & \overline{x}_{k+1,i} +  K_{1,k+1,i} b_{k+1,i},  \notag
\end{align}
where
\begin{align}
  \overline{x}_{k+1,i}  = &  \hat{f}(\hat{x}_{k,i}) + a \sum_{j \in \mathcal{N}_i} \pi_{ij} G  \hat{x}_{k,j}, \notag  \\
  b_{k+1,i} = &  y_{k+1,i} - H_{i}  \overline{x}_{k+1,i} , \notag \\
  K_{1,k+1,i} = &  ( \breve{P}_{k+1,i}^{-1}  + H_{i}^{T} \hat{R}_{k+1,i}^{-1} H_{i} +  \overline{\lambda}_{k,i} I_n  )^{-1} H_{i}^{T}  \hat{R}_{k+1,i}^{-1}, \notag
\end{align}
where $P_{k+1,i}$, $\breve{P}_{k+1,i}$, $\hat{R}_{k+1,i}$, $\overline{\lambda}_{k,i}$, and $\alpha_{k,i}$ are defined as those in (\ref{equ:J22}).

Then, denote the priori and posterior estimation errors for node $i$ in the distributed estimator (\ref{equ:estimate3}) as  $\overline{e}_{k,i} = x_{k,i} - \overline{x}_{k,i} $ and $e_{k,i} = x_{k,i} - \hat{x}_{k,i} $, respectively. It follows from the above equations that
\begin{align}
 e_{k+1,i} =  &  [I_n -  K_{1,k+1,i} H_{i}] \overline{e}_{k+1,i}   - K_{1,k+1,i} \nu_{k+1,i}, \notag
\end{align}
and
\begin{align}
 \overline{e}_{k+1,i}   = & (F_{i} + a \pi_{ii} G + E_{2,i} \Delta_{2,k,i} ) e_{k,i}   + a \sum_{j \in \mathcal{N}_i, j \neq i} \pi_{ij} G  e_{k,j}  \notag \\
 & + \omega_{k,i}  . \notag
\end{align}

Subsequently, if Assumption \ref{asm1} holds, the dynamics of the augmented state error $e_{k+1}=[e_{k+1,1}^T$, $\ldots$, $e_{k+1,N}^T]^T$ can be derived as
\begin{align} \label{equ:ee}
 e_{k+1} =  &   \mathcal{H}_{k+1}  F \Gamma_k  e_{k} - K_{1,k+1} \nu_{k+1} + \mathcal{H}_{k+1}   \omega_{k},
\end{align}
where $\mathcal{H}_{k+1} = I_{Nn} -  K_{1,k+1} H$,
$F = \text{diag} [F_{1}$, $\ldots$, $F_{N}]$,
$H = \text{diag} [H_{1}$, $\ldots$, $H_{N}]$,
\ $\Gamma_k = I_{Nn} + F^{-1} E_{2} \Delta_{2,k} + a F^{-1} \overline{G}$, $K_{1,k+1} = \text{diag} [K_{1,k+1,1}$, $\ldots$, $K_{1,k+1,N}]$, $\nu_{k+1}^T=[\nu_{k+1,1}^T$, $\ldots$, $\nu_{k+1,N}^T]^T$,
$E_{2} = \text{diag} [E_{2,1}$, $\ldots$, $E_{2,N}]$,
$\Delta_{2,k} = \text{diag} [\Delta_{2,k,1}$, $\ldots$, $\Delta_{2,k,N}]$,
$\omega_{k}^T=[\omega_{k,1}^T$, $\ldots$, $\omega_{k,N}^T]^T$,
$\overline{G}=[{G^{1}}^T$, $\ldots$, ${G^{N}}^T]^T$ with $G^{i} = [\pi_{i1} G$, $\ldots$, $\pi_{i(i-1)} G$, $0$, $\pi_{i(i+1)} G$, $\ldots$, $\pi_{iN} G]$,  and other matrices are given in (\ref{equ:PP}).

As guaranteed by Theorem \ref{thm4}, the parameters of estimator (\ref{equ:estimate3}) are bounded at every step $k$, i.e., $K_{1,k+1}$ and $\mathcal{H}_{k+1}$ in (\ref{equ:ee}) are uniformly bounded. To ensure that the mean-squared error, i.e., $e_{k+1}^T e_{k+1}$, is bounded, it suffices to guarantee that $\mathcal{H}_{k+1}  F \Gamma_k $ in (\ref{equ:ee}) be  ultimately stable, or sufficiently, $\sigma (\mathcal{H}_{k+1} F \Gamma_k)$ is less than one.
As presented in Appendix \ref{proof4} and based on the existing results on Riccatti recursions \cite{zhou1996robust}, as $k \to \infty$, one has
\begin{align} \label{equ:mathcal_H}
\mathcal{H}_{k+1}  F = \mathcal{H}  F =  (I_{Nn}  +  (\breve{P}^{-1}   + \overline{\lambda}  )^{-1} H^{T} \hat{R}^{-1} H )^{-1} F,
\end{align}
where $\mathcal{H}  F$ is stable and $  \breve{P} =  (1+a) F (\breve{P}^{-1}  + H^{T} \hat{R}^{-1} H + \overline{\lambda} )^{-1} F  + Q $.
Thus, as $k \to \infty$, one has
\begin{align}
 \sigma_{max} (\mathcal{H}_{k+1}  F \Gamma_k)^2 = & \max \{ \lambda  ( \mathcal{H}  F \Gamma_k \Gamma_k^T F^T \mathcal{H}^T ) \} \notag \\
= & \| \mathcal{H}  F \Gamma_k \Gamma_k^T F^T \mathcal{H}^T \|_2  \notag \\
\leq & \| \mathcal{H}  F \|_2^2  \| \Gamma_k \|_2^2.  \notag
\end{align}

Based on Assumption \ref{asm3}, if (\ref{equ:condition_bound}) holds, one has
\begin{align}
 \sigma_{max} (\mathcal{H}_{k+1}  F \Gamma_k) < 1 . \notag
\end{align}

Thus, the proof of Theorem \ref{thm5} is complete.

\bibliographystyle{IEEEtran}
\bibliography{ref}

\begin{thebibliography}{10}
\providecommand{\url}[1]{#1}
\csname url@samestyle\endcsname
\providecommand{\newblock}{\relax}
\providecommand{\bibinfo}[2]{#2}
\providecommand{\BIBentrySTDinterwordspacing}{\spaceskip=0pt\relax}
\providecommand{\BIBentryALTinterwordstretchfactor}{4}
\providecommand{\BIBentryALTinterwordspacing}{\spaceskip=\fontdimen2\font plus
\BIBentryALTinterwordstretchfactor\fontdimen3\font minus
  \fontdimen4\font\relax}
\providecommand{\BIBforeignlanguage}[2]{{%
\expandafter\ifx\csname l@#1\endcsname\relax
\typeout{** WARNING: IEEEtran.bst: No hyphenation pattern has been}%
\typeout{** loaded for the language `#1'. Using the pattern for}%
\typeout{** the default language instead.}%
\else
\language=\csname l@#1\endcsname
\fi
#2}}
\providecommand{\BIBdecl}{\relax}
\BIBdecl

\bibitem{Arenas2008}
A.~Arenas, A.~Diaz-Guilera, J.~Kurths, Y.~Moreno, and C.~Zhou,
  ``\BIBforeignlanguage{English}{Synchronization in complex networks},''
  \emph{\BIBforeignlanguage{English}{Physics Reports}}, vol. 469, no.~3, pp.
  93--153, 2008.

\bibitem{wen2019pinning}
G.~{Wen}, P.~{Wang}, X.~{Yu}, W.~{Yu}, and J.~{Cao}, ``Pinning synchronization
  of complex switching networks with a leader of nonzero control inputs,''
  \emph{IEEE Transactions on Circuits and Systems I: Regular Papers}, vol.~66,
  no.~8, pp. 3100--3112, Aug 2019.

\bibitem{duanpeihu2019}
P.~Duan, Z.~Duan, Y.~Lv, and G.~Chen,
  ``\BIBforeignlanguage{English}{Distributed finite-horizon extended {K}alman
  filtering for uncertain nonlinear systems},''
  \emph{\BIBforeignlanguage{English}{IEEE Transactions on Cybernetics}}, in
  press, 2019, doi: 10.1109/TCYB.2019.2919919.

\bibitem{SHI2016275}
D.~Shi, T.~Chen, and M.~Darouach, ``Event-based state estimation of linear
  dynamic systems with unknown exogenous inputs,'' \emph{Automatica}, vol.~69,
  pp. 275 -- 288, 2016.

\bibitem{duanpeihu2020}
P.~{Duan}, G.~{Lv}, Z.~{Duan}, and Y.~{Lv}, ``Resilient state estimation for
  complex dynamic networks with system model perturbation,'' \emph{IEEE
  Transactions on Control of Network Systems}, in press, 2020, doi:
  10.1109/TCNS.2020.3035759.

\bibitem{shen2011bounded}
B.~{Shen}, Z.~{Wang}, and X.~{Liu}, ``Bounded {$H_{\infty}$} synchronization
  and state estimation for discrete time-varying stochastic complex networks
  over a finite horizon,'' \emph{IEEE Transactions on Neural Networks},
  vol.~22, no.~1, pp. 145--157, Jan 2011.

\bibitem{Ding2012H}
D.~Ding, Z.~Wang, B.~Shen, and H.~Shu,
  ``\BIBforeignlanguage{English}{${H}_{\infty}$ state estimation for
  discrete-time complex networks with randomly occurring sensor saturations and
  randomly varying sensor delays},'' \emph{\BIBforeignlanguage{English}{IEEE
  Transactions on Neural Networks and Learning Systems}}, vol.~23, no.~5, pp.
  725--736, 2012.

\bibitem{shen2013H}
B.~{Shen}, Z.~{Wang}, D.~{Ding}, and H.~{Shu}, ``{$H_{\infty}$} state
  estimation for complex networks with uncertain inner coupling and incomplete
  measurements,'' \emph{IEEE Transactions on Neural Networks and Learning
  Systems}, vol.~24, no.~12, pp. 2027--2037, Dec 2013.

\bibitem{wang2016anevent}
L.~Wang, Z.~Wang, T.~Huang, and G.~Wei, ``\BIBforeignlanguage{English}{An
  event-triggered approach to state estimation for a class of complex networks
  with mixed time delays and nonlinearities},''
  \emph{\BIBforeignlanguage{English}{IEEE Transactions on Cybernetics}},
  vol.~46, no.~11, pp. 2497--2508, 2016.

\bibitem{Yong2017asy}
Y.~Xu, R.~Lu, H.~Peng, K.~Xie, and A.~Xue,
  ``\BIBforeignlanguage{English}{Asynchronous dissipative state estimation for
  stochastic complex networks with quantized jumping coupling and uncertain
  measurements},'' \emph{\BIBforeignlanguage{English}{IEEE Transactions on
  Neural Networks and Learning Systems}}, vol.~28, no.~2, pp. 268--277, 2017.

\bibitem{Wu2018state}
X.~Wu, G.-P. Jiang, and X.~Wang, ``\BIBforeignlanguage{English}{State
  estimation for general complex dynamical networks with packet loss},''
  \emph{\BIBforeignlanguage{English}{IEEE Transactions on Circuits and Systems
  II: Express Briefs}}, vol.~65, no.~11, pp. 1753--1757, 2018.

\bibitem{dong2018variance}
H.~{Dong}, N.~{Hou}, Z.~{Wang}, and W.~{Ren}, ``Variance-constrained state
  estimation for complex networks with randomly varying topologies,''
  \emph{IEEE Transactions on Neural Networks and Learning Systems}, vol.~29,
  no.~7, pp. 2757--2768, July 2018.

\bibitem{hu2013recursive}
J.~Hu, Z.~Wang, and H.~Gao, ``Recursive filtering with random parameter
  matrices, multiple fading measurements and correlated noises,''
  \emph{Automatica}, vol.~49, no.~11, pp. 3440--3448, 2013.

\bibitem{HU2016A}
J.~Hu, Z.~Wang, S.~Liu, and H.~Gao, ``A variance-constrained approach to
  recursive state estimation for time-varying complex networks with missing
  measurements,'' \emph{Automatica}, vol.~64, pp. 155 -- 162, 2016.

\bibitem{hu2019variance}
J.~{Hu}, Z.~{Wang}, G.~{Liu}, and H.~{Zhang}, ``Variance-constrained recursive
  state estimation for time-varying complex networks with quantized
  measurements and uncertain inner coupling,'' \emph{IEEE Transactions on
  Neural Networks and Learning Systems}, vol.~31, no.~6, pp. 1955--1967, 2019.

\bibitem{li2017nonaugmented}
W.~Li, Y.~Jia, and J.~Du, ``Non-augmented state estimation for nonlinear
  stochastic coupling networks,'' \emph{Automatica}, vol.~78, pp. 119--122,
  2017.

\bibitem{li2017state}
W.~{Li}, Y.~Jia, and J.~Du, ``State estimation for stochastic complex networks
  with switching topology,'' \emph{IEEE Transactions on Automatic Control},
  vol.~62, no.~12, pp. 6377--6384, 2017.

\bibitem{li2019resilient}
W.~{Li}, Y.~{Jia}, and J.~{Du}, ``Resilient filtering for nonlinear complex
  networks with multiplicative noise,'' \emph{IEEE Transactions on Automatic
  Control}, vol.~64, no.~6, pp. 2522--2528, June 2019.

\bibitem{Zhou2013}
T.~{Zhou}, ``Coordinated one-step optimal distributed state prediction for a
  networked dynamical system,'' \emph{IEEE Transactions on Automatic Control},
  vol.~58, no.~11, pp. 2756--2771, Nov 2013.

\bibitem{SAYED2002A}
A.~H. Sayed, V.~H. Nascimento, and F.~A.~M. Cipparrone, ``A regularized robust
  design criterion for uncertain data,'' \emph{SIAM Journal on Matrix Analysis
  \& Applications}, vol.~23, no.~4, pp. 1120--1142, 2002.

\bibitem{zhou1996robust}
K.~Zhou, J.~C. Doyle, K.~Glover \emph{et~al.}, \emph{Robust and {O}ptimal
  {C}ontrol}.\hskip 1em plus 0.5em minus 0.4em\relax Prentice Hall, 1996.

\bibitem{xiong2012robust}
K.~Xiong, C.~Wei, and L.~Liu, ``Robust {Kalman} filtering for discrete-time
  nonlinear systems with parameter uncertainties,'' \emph{Aerospace Science and
  Technology}, vol.~18, no.~1, pp. 15--24, 2012.

\bibitem{xiao2016tracking}
B.~{Xiao}, S.~{Yin}, and O.~{Kaynak}, ``Tracking control of robotic
  manipulators with uncertain kinematics and dynamics,'' \emph{IEEE
  Transactions on Industrial Electronics}, vol.~63, no.~10, pp. 6439--6449, Oct
  2016.

\bibitem{theodor1996robust}
Y.~Theodor and U.~Shaked, ``Robust discrete-time minimum-variance filtering,''
  \emph{IEEE Transactions on Signal Processing}, vol.~44, no.~2, pp. 181--189,
  1996.

\bibitem{arora2009computational}
S.~Arora and B.~Barak, \emph{{Computational Complexity: a Modern
  Approach}}.\hskip 1em plus 0.5em minus 0.4em\relax Cambridge University
  Press, 2009.

\bibitem{BartholomewBiggs2008Nonlinear}
M.~Bartholomew-Biggs, \emph{Nonlinear Optimization with Engineering
  Applications}.\hskip 1em plus 0.5em minus 0.4em\relax Springer Science \&
  Business Media, 2008.

\bibitem{convexoptimization}
S.~P. Boyd and L.~Vandenberghe, \emph{\BIBforeignlanguage{English}{Convex
  Optimization}}.\hskip 1em plus 0.5em minus 0.4em\relax Cambridge University
  Press, 2004.

\bibitem{sayed2001framework}
A.~H. Sayed, ``A framework for state-space estimation with uncertain models,''
  \emph{IEEE Transactions on Automatic Control}, vol.~46, no.~7, pp. 998--1013,
  2001.

\bibitem{zhou2010on}
T.~{Zhou}, ``On the convergence and stability of a robust state estimator,''
  \emph{IEEE Transactions on Automatic Control}, vol.~55, no.~3, pp. 708--714,
  March 2010.

\bibitem{BATTISTELLI201875}
G.~Battistelli, L.~Chisci, and D.~Selvi, ``A distributed {K}alman filter with
  event-triggered communication and guaranteed stability,'' \emph{Automatica},
  vol.~93, pp. 75 -- 82, 2018.

\bibitem{Bonnabel2015}
S.~{Bonnabel} and J.~{Slotine}, ``A contraction theory-based analysis of the
  stability of the deterministic extended {K}alman filter,'' \emph{IEEE
  Transactions on Automatic Control}, vol.~60, no.~2, pp. 565--569, Feb 2015.

\end{thebibliography}

\end{document}